\documentclass{article}

\usepackage{graphicx}
\usepackage{amssymb}
\usepackage{booktabs}
\usepackage{color}
\usepackage{amsmath}
\usepackage{epsfig,amsthm}
\usepackage{float}
\usepackage{hyperref}
\usepackage{caption}

\usepackage{arxiv}

\usepackage[utf8]{inputenc} 
\usepackage[T1]{fontenc}    
\usepackage{hyperref}       
\usepackage{url}            
\usepackage{booktabs}       
\usepackage{amsfonts}       
\usepackage{nicefrac}       
\usepackage{microtype}      
\usepackage{lipsum}		
\usepackage{graphicx}
\usepackage{natbib}
\usepackage{doi}

\title{Integrating Tick-level Data and Periodical Signal for High-frequency Market Making}

\author{ Jiafa He \\
        \texttt{jhebu@connect.ust.hk}\\
        Department of Mathematics\\
	The Hong Kong University of Science and Technology\\
        \And Cong ZHENG\\
        \texttt{czhengae@connect.ust.hk}\\
	Department of Mathematics\\
	The Hong Kong University of Science and Technology\\
        \And Can YANG\\
        \texttt{macyang@ust.hk}\\
	Department of Mathematics\\
	The Hong Kong University of Science and Technology\\
}

\date{}

\begin{document}
\maketitle

\begin{abstract}
we focus on the problem of market making in high-frequency trading. Market making is a critical function in financial markets that involves providing liquidity by buying and selling assets. However, the increasing complexity of financial markets and the high volume of data generated by tick-level trading makes it challenging to develop effective market making strategies. To address this challenge, we propose a deep reinforcement learning approach that fuses tick-level data with periodic prediction signals to develop a more accurate and robust market making strategy. Our results of market making strategies based on different deep reinforcement learning algorithms under the simulation scenarios and real data experiments in the cryptocurrency markets show that the proposed framework outperforms existing methods in terms of profitability and risk management.
\end{abstract}


\section{Introduction}
\subsection{Background}
The widespread use of the electronic limit order book (LOB) in modern financial markets has led to the rapid growth of high-frequency trading strategies, which now account for over 65\% of the total trading volume in financial markets\cite{hagstromer2013diversity}. Market Making is one of the most common strategies in high-frequency trading, where an agent, called a market maker, simultaneously and continually posts both bid limit orders and ask limit orders on the limit order book of a specific security. As liquidity providers, market making is an important participant in markets to help to reduce short-term volatility and transaction cost. The goal of market making is to make a profit from the quoted spread, the difference between the price of bid orders and ask orders. However, market makers face different types of risks when pursuing profit. The first type of risk is called inventory risk, which occurs when a market maker unintentionally accumulates a nonzero inventory due to unfavorable price moving of an underlying asset. For example, if the true fair price of the asset is \$90 and a market maker estimates the fair price to be between \$95 and \$105, then quotes a bid order at \$90 and an ask order at \$105, only the bid order may be executed, potentially leading to a positive inventory. The second type of risk is adverse selection risk, which stems from information asymmetry. Since market makers' limit orders need to be quoted in the LOB waiting to be executed, the information on these limit orders is exposed to all other participants in the markets. Then, speculation traders can take advantage of the information and make a profit from market makers, especially those who are slow in reacting to price volatility. The other types of risks include latency and model uncertainty risks. Therefore, in order to maximize risk-adjusted returns, an optimal market making agent needs to dynamically adjust the price and size of the bid and ask limit orders based on the information from both markets and the agent.

\subsection{Related Work}

Most commonly, there are two main branches for solving the optimal market making problem. The first branch is analytical approaches. Based on market making's inventory control mechanism, the seminal work \cite{avellaneda2008high} proposes an analytical market making framework, the Avellaneda-Stoikov (AS) model, with a closed-form approximation of the optimal quotes by solving the Hamilton-Jacobi-Bellman (HJB) equations. Following a similar recipe, many variants of the AS model are developed. To be more realistic, \cite{gueant2013dealing} takes inventory limits into consideration and shows the asymptotic behavior of the optimal quotes.
Furthermore, \cite{fodra2012high} extend the models in \cite{avellaneda2008high} and \cite{gueant2013dealing} to a general non-martingale framework of mid-price processes. To include additional features, \cite{guilbaud2013optimal} extracts information from the bid-ask spread by estimating the transition matrix and intensity parameters for the spread. \cite{cartea2018algorithmic} considers the features from the shape of the LOB and the impact of market orders. \cite{cartea2017algorithmic} derives a robust optimal market making by incorporating ambiguity or uncertainty aversion in the market making's choice leading to a reduction in the standard deviation of profits. Typically, analytical approaches share several common weaknesses. Firstly, a series of strong assumptions are essential in such models. Secondly, historical market data is needed to calibrate the environmental parameters. Lastly, the complex behavior of the LOB and the fat-tail characteristics of the high-frequency market data are not considered in most of the analytical approaches. 

The second branch is data-driven approaches, which learn directly from real market data. Hence, these learning-based approaches are model-free and more suitable for real-world market making strategies than the analytical methods. Since optimal market making is naturally formalized as a Markov decision process (MDP), a discrete-time stochastic control process, reinforcement learning (RL), as a popular machine learning technique to solve the MDPs in decision making, begins to be applied to the optimal market making problems in the literature. In the early stage, The first work, utilizing reinforcement learning methods to solve market making problems, is proposed in \cite{chan2001electronic}, in which a market making agent is designed to learn from market experiences and maximize its profits without any inside knowledge of the market environment. Later \cite{kim2002modeling} improves the above method by embedding the information from order flow and the dynamics of the LOB into the RL-based model. Though the RL models proposed in \cite{chan2001electronic} and \cite{kim2002modeling} are relatively simple, these approaches light the way for RL applications in more complex optimal market making problems. 

Currently, motivated by the above methods, various temporal-difference reinforcement learning models are applied in the literature. \cite{lim2018reinforcement} shows a novel use of the CARA utility to improve an optimal policy trained using Q-learning. \cite{spooner2018market} uses a linear combination of tile codings as a value function approximator in a temporal-difference reinforcement learning, called SARSA. \cite{haider2019effect} takes into account the impact of the spread over the market makers in the reward and obtains a significant improvement in the magnitude of returns. \cite{zhong2020data} designs a simple lookup table based on off-policy Q-learning and outperforms the benchmark strategies in out-of-sample testing.

Recently, due to the high complexity and non-linearity of the optimal market making strategy, deep reinforcement learning (DRL) is finding wider and wider applications in this field, which utilizes deep learning models to implement policies representation and state-action function approximation. Compared to non-deep reinforcement learning methods, DRL models is suitable for more complex scenarios. For example, \cite{sadighian2019deep} implements market making strategies based on two advanced policy gradient-based DRL algorithms and the DRL model is trained on a simulator based on realistic LOB data. \cite{gueant2019deep} uses a model-based deep actor-critic algorithm to solve a multi-asset market making in corporate bond markets in high dimension. \cite{patel2018optimizing} proposes a multi-agent reinforcement learning framework, where a macro-agent model is in charge of direction decision making, such as buy, sell or hold and another micro-agent model is used to optimize the quoted price of the bid or ask limit order.

However, for the convenience of the environment and agent modeling and feature engineering, most RL-based approaches reshape the real event-driven market data, also called tick-level market data, to a periodical data to extract predictive signals and train the market making agent, which causes a loss of information in real market data. Moreover, they tend to simplify the order routing process and disregard the latency of the network in practice, which plays a crucial role in high-frequency trading strategies. Therefore, the goal of this work is to contribute to these subjects in a real-world application, including merging tick-level data information with periodical data information and taking latency into consideration.

\subsection{Our Work And Contributions}
In this work, the main target is to design a unified reinforcement learning framework to combine information from tick-level LOB data and periodical data for market making strategies with latency. For this purpose, we first reconstruct a local full limit order book based on three types of real event-driven market data from cryptocurrency markets, such as limit order placing, limit order canceling and limit order execution. Then, we develop a deep reinforcement learning model with a tick-level environment and a latency-aware agent to learn the tick-level information from the event-driven limit order book data. Since the tick-level data is the basic element in the market trading environment, which directly reflects instantaneous supply and demand, the tick-level DRL model can promptly respond to the change in price and also precisely control the inventory risk. Moreover, in order to reduce the adverse selection risk, a feature extraction framework is designed to learn the predictive information from the periodical market data. Finally, we embed the information from both tick-level data and periodical data into the training phase of the DRL model to improve the performance of our market making strategy. Our contributions to this work are the following:

First, we design a tick-level state and action space in our basic deep reinforcement learning model. Specifically, contrary to most traditional approaches where the LOB is updated on a fixed periodical time interval, the proposed tick-level DRL model updates the LOB data and the agent states in a tick-by-tick way, which provides more timely information about the change of the market to the agent. The most related work is \cite{gavsperov2021market}, where a reinforcement learning model is proposed also based on the tick data. However, the tick-level market data is fed based on a non-physical time in \cite{gavsperov2021market}, meaning that the timestamp of the tick data is ignored in this model. In our basic framework, we not only take advantage of the promptness of the tick data but also take into account the physical time information, which paves the way for the further expansion of the basic framework, such as the predictive signal embedding and latency analysis. Additionally, we verify the tick-level model in different simulation scenarios, which outperforms the baseline models in regard to a series of metrics.

Second, we analyze the predictability of various features in real cryptocurrency market data based on the fixed periodical time interval. Inspired by the work in the traditional market \cite{ait2022and}, we construct the high-frequency predictors from trades and quote data and examine the predictability of these features based on the future high-frequency returns and durations.

Third, since our basic DRL model is a tick-level physical time framework, we can easily embed the periodical predictive signal into the unified model. We verify the improvement of the fused model in the real data scenario and achieve better performance over baseline models.

Finally, Considering that the latency, stemming from the order routing process, plays a key role in the profitability of market makers in reality, we conduct a comprehensive latency analysis based on our basic tick-level model. The related work is \cite{gao2020optimal}, which studies the optimal market making for large-tick assets in the presence of latency and shows negative impacts of the latency on the performance. However, \cite{gao2020optimal} only verifies results based on periodical data. To the authors' knowledge, this is the first study to conduct the analysis of the latency in the tick data scenario based on the DRL model. We separate the latency into quoting order latency and canceling order latency and analyze the impacts of each type of latency on the performance of market making model.

The proposed unified framework that consists of a tick-level agent for short-term ultra-high-frequency information and a predictor for long-term periodical information as shown in Figure \ref{drl_mm_framework}. The rest of the paper is organized as follows: Section \ref{Market_Data_Pre} describes the collection of raw market data and the data preprocessing required to reconstruct the limit order book (LOB) for further analysis. In section \ref{Agent_Learning}, we construct a simulated tick-level environment based on the top order book data and design a reinforcement learning (RL)-based market making agent to learn from the tick-level environment. The simulation results indicate that the RL-based agent can achieve higher performance in risk-adjusted PnL than baseline agents. Section \ref{Feature_Extraction} focuses on feature extraction questions, specifically on how to design and evaluate features with the periodical LOB data to predict future returns or price movements. In section \ref{Assembled_Framework}, we use a unified framework to combine the tick-level signal and the periodical signal into the RL-based agent. The results of real data experiments demonstrate a significant improvement over the benchmark agents.

\section{Market Data Preprocessing}\label{Market_Data_Pre}

In this work, our main focus is on the cryptocurrency markets. These markets can be broadly categorized into two main components - decentralized exchanges (DEXs) and centralized exchanges (CEXs).
Decentralized exchanges operate using trading engines that are built on the blockchain and rely primarily on an auto market-making system that is meant to ensure a stable supply of liquidity pools. On the other hand, centralized exchanges are modeled on the traditional financial markets concept of a central limit order book. This book aggregates all bid and ask limit orders together into a public order book to be matched with market orders.
Currently, the CEXs dominate the average trading volumes in the crypto market ecosystem. In fact, on average, the trading volumes on CEXs are more than ten times that of DEXs. As such, in this work, we've decided to focus primarily on CEXs with the aim of capturing a significant fraction of the globally traded volume.
The most popular CEXs in the crypto market are Binance, Okex, Coinbase, and Kraken, with Coinbase being the most widely-used platform. One of the reasons for this is that Coinbase provides level-3 information on market data. This information is crucial because it allows us to track the status of every limit order during its life cycle in the market. It is for this reason that we've chosen to study the market data from Coinbase in the follow-up analysis.


We gather level-3 market data from Coinbase using a multi-step process. Our first step is to receive live tick-level stream data by connecting to the exchange's websocket API endpoints. These endpoints offer information about limit orders and trades. Specifically, for trade data, the websocket API endpoints publish information about one trade to every connector immediately after one limit order in the book is matched with one market order. For limit order data, the exchange publishes new data about one limit order whenever there is a status change in the limit order. It is worth noting that there are different types of limit orders, such as buy limit orders and sell limit orders, and each type has its own unique characteristics. Therefore, it is important to have accurate information about the status change of each limit order to gain a comprehensive understanding of market trends. In order to maintain an up-to-date local level-3 order book, we are also required to request a full order book snapshot from the REST feed. This snapshot provides a full picture of the current state of the order book. We then use the real-time stream messages from the websocket channels to update the order book snapshot. This ensures that we have the most current view of the market. To make offline market analysis more convenient, we store both the snapshot data and the live stream data from the websocket endpoints in our local database. This enables us to perform detailed market analysis and draw insights that can help us make informed decisions. Overall, our multi-step approach to collecting level-3 market data from Coinbase allows us to stay up-to-date with the latest market trends and make informed decisions based on accurate and comprehensive information.


We provide a detailed explanation of the process involved in reconstructing an order book after data collection. First, the local order book is initialized with the order book snapshot from the REST feed. Next, the stream messages with sequence numbers that are smaller or equal to the snapshot sequence number are discarded, and only the remaining messages are used to update the order book. The next step involves identifying different types of stream messages and updating the local order book accordingly. If the message type is 'received,' it indicates that a new incoming order has been accepted by the exchange's matching engine for processing and has not been listed on the book. In this case, the local order book remains unchanged. On the other hand, if the message type is 'open,' it means that the limit order is now open on the order book. In this case, the local order book must add the limit order to the corresponding price level. Similarly, if the message type is 'done,' it means that the order is no longer on the book due to being canceled or filled. In this instance, the local order book needs to remove this order from the list. Finally, if the message type is 'match,' it indicates that a trade occurred between a limit order and a market order. In summary, the process of reconstructing the order book after data collection involves initializing the local order book with the order book snapshot, discarding irrelevant stream messages, and updating the order book based on different message types. This process ensures that the local order book stays up-to-date and reflects the latest market activity.

\begin{figure*}
  \begin{center}
  \includegraphics[width=0.5\linewidth]{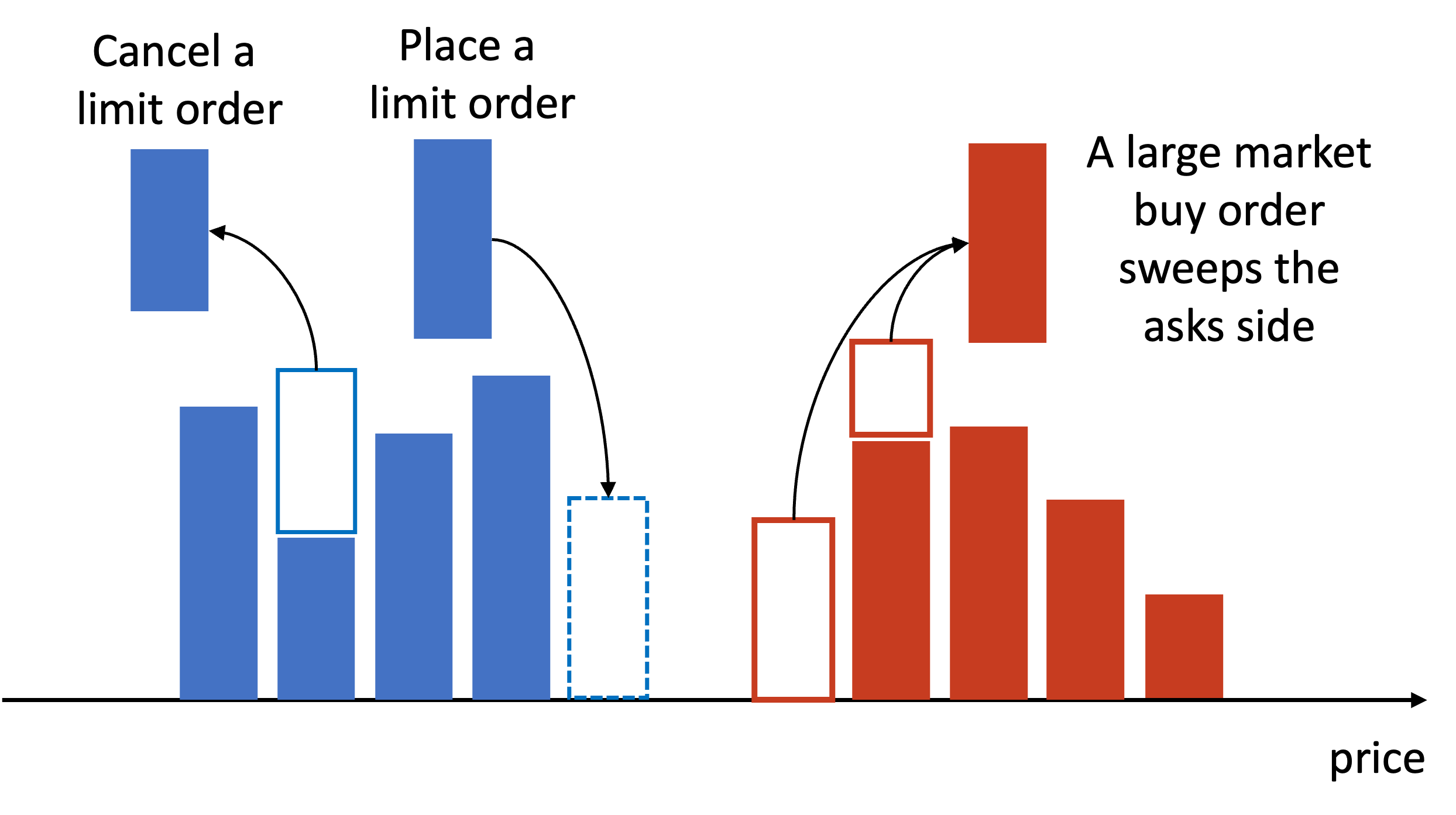}
  \end{center}
    \caption{\textbf{Limit order book reconstruction}. In the LOB, the bid side is displayed in a blue box and consists of limit orders placed by traders who wish to purchase the asset at a specified price. On the other hand, the ask side is shown in a red box and contains limit orders from traders who wish to sell the asset at a particular price. By viewing the LOB, traders can evaluate the current supply and demand for the asset and make informed trading decisions.}
  \label{drl_mm_limit_order_book}
\end{figure*}

\section{Tick-level Agent Learning}\label{Agent_Learning}

The key question for optimal market making is how to dynamically adjust the agent's quoting orders to maximize profit while controlling risks. Reinforcement learning models provide a natural learning-based framework for such a decision-making process. Therefore, our objective in this section is to develop a naive tick-level reinforcement learning agent that relies on top limit order book data and mainly focuses on inventory risk. To facilitate analysis, we first create a simulation model based on the characteristics of the real top LOB data. Next, we provide details about the proposed reinforcement learning framework. Finally, we conduct various experiments on simulations that demonstrate the effectiveness of the RL-based framework over baseline models. In later sections, we enhance the naive framework by incorporating additional LOB microstructural features to reduce the risk of adverse selection.

\subsection{Top Limit Order Book Simulation}

It is well established in the literature that the top limit order book, which consists of the top bid and ask levels, is one of the most important parts of LOB data \cite{bouchaud2002statistical}\cite{cao2009information}\cite{cont2014price}. Therefore, we begin building our naive RL-based model by analyzing the dynamics of the top limit order book data first. 

We discuss the main elements in the top limit order book data. Firstly, the basic elements of the top order book are the best bid price $p^b$ and the best ask price $p^a$, which represent the highest price the bid side is willing to offer to buy the asset and the lowest price the ask side is willing to offer to sell the asset, respectively. Furthermore, the distance between the best bid price and the best ask price is called the spread, denoted by $s = (p^a - p^b)/p^b$, which is a crucial metric as it indicates the liquidity of the market. A smaller spread signifies a more liquid market, which means there is a higher likelihood of executing a trade at the desired price. On the other hand, a wider spread implies a less liquid market, which could lead to difficulty in executing trades without affecting the market price. Finally, the middle price is the average of the best bid price and the best ask price, denoted by $p^{mid}=(p^{b}+p^{a})/2$. It is commonly used as an indicator to predict the true price, also known as the fair price.

\begin{figure*}
  \begin{center}
  \includegraphics[width=0.5\linewidth]{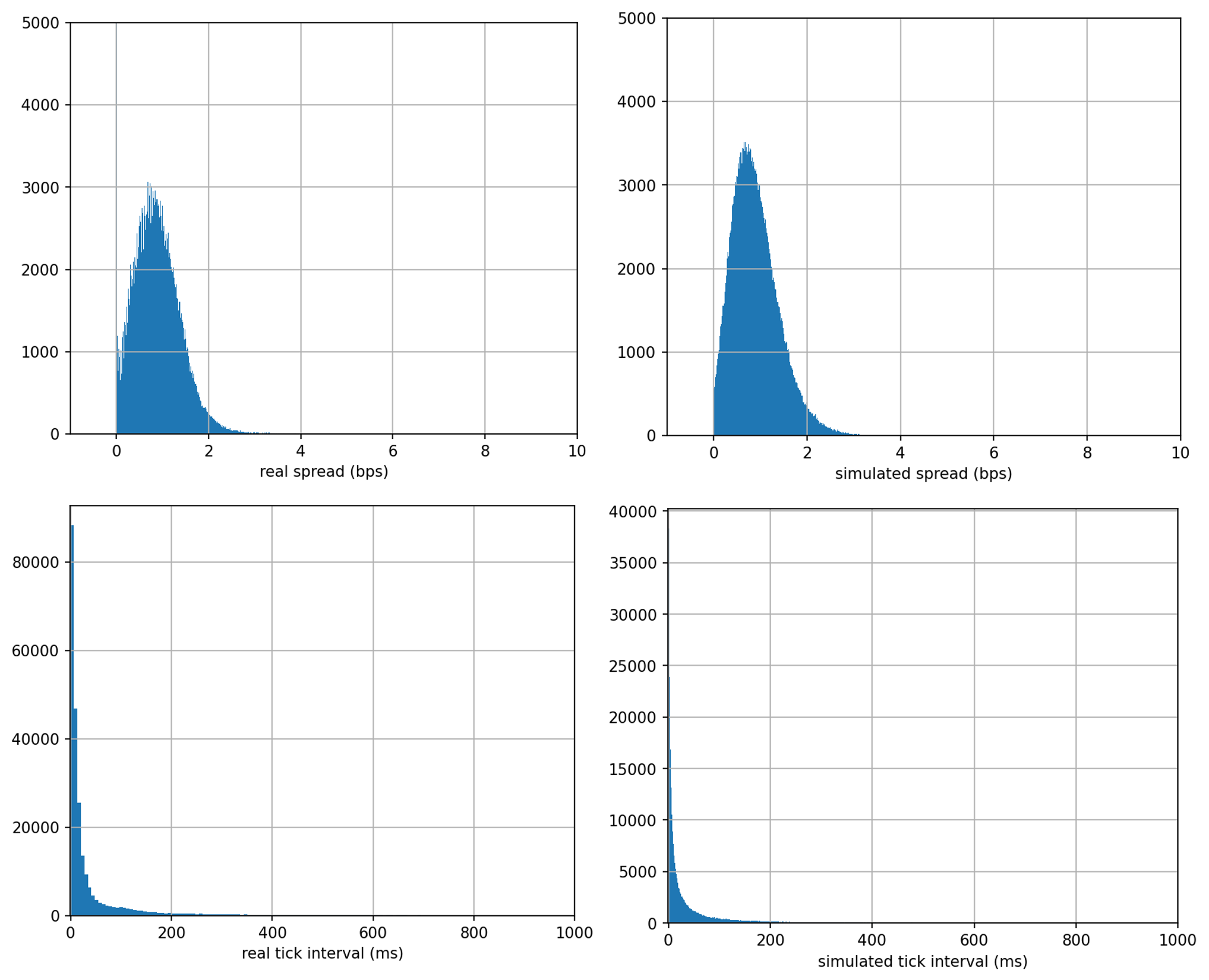}
  \end{center}
    \caption{\textbf{Histograms of spread and tick interval time for real data and simulated data}. The first row compares the distribution of real and simulated ask-bid spreads. The second row compares the time intervals of two adjacent tick data between real data and simulated data.}
  \label{drl_mm_simulation_data_hist}
\end{figure*}

To facilitate analysis and comparison, we build a market model to simulate the top limit order book, which is calibrated with the real top order book data we reconstructed in section \ref{Market_Data_Pre}. The details of the simulation model are discussed as follows: We first model the time interval between two ticks as a log-normal distribution, denoted as 
\begin{equation}
    \Delta T =e^{\mu+\sigma Z}
\end{equation}
where $\mu$ and $\sigma$ are the expected value and standard deviation of the time interval $\Delta T$'s natural logarithm and $Z$ is a standard normal variable. Then, based on the time interval model, we assume the middle price $p^{mid}$ as a random walk model, which is given by
\begin{equation}
    p^{mid}_{n} = p^{mid}_{n-1} \times \left(1+ \Delta P \times \Delta T \right)
\end{equation}
where $p^{mid}_{n}$ and $p^{mid}_{n-1}$ represent the middle price of the $n^{th}$ and $(n-1)^{th}$ tick data, $\Delta T$ is the time interval defined above and $\Delta P$ indicates the price change standard deviation, which follows a zero mean normal distribution, that is $\Delta P \sim \mathcal{N}\left(0, \sigma^2\right)$. Finally, we also model the spread as a log-normal distribution, as follows:
\begin{equation}
    s =e^{\mu+\sigma Z}
\end{equation}
where similarly $\mu$ and $\sigma$ are the expected value and standard deviation of the spread $s$'s natural logarithm and $Z$ is a standard normal variable. Based on above assumption, the simulated best bid price and best ask price is given as:
\begin{equation}
\begin{split}
    p^{best\_bid}_{n} &= p^{mid}_{n} \times (1-s/2) \\
    p^{best\_ask}_{n} &= p^{mid}_{n} \times (1+s/2)
\end{split}
\end{equation}

\begin{figure*}
  \begin{center}
  \includegraphics[width=0.5\linewidth]{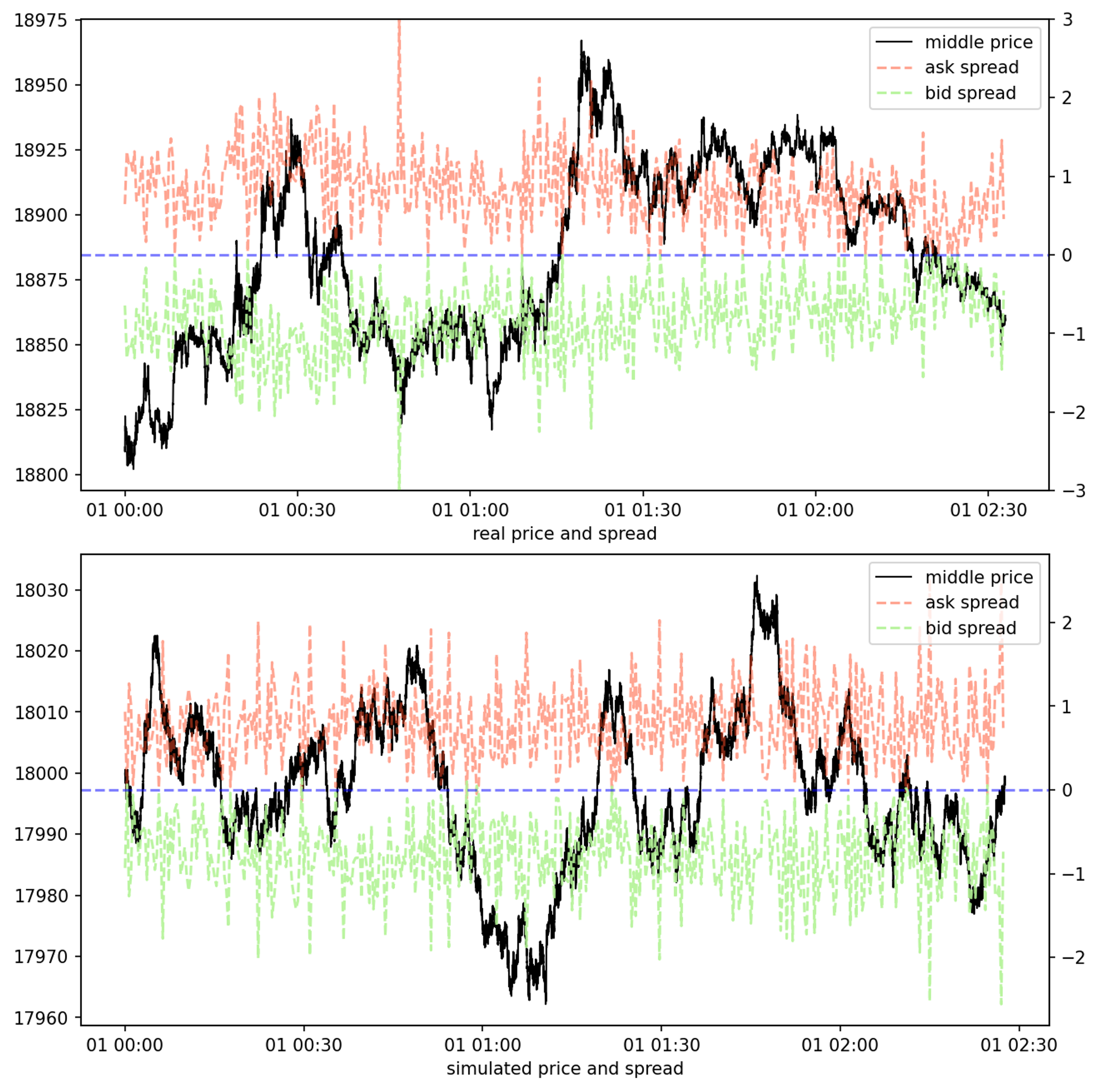}
  \end{center}
    \caption{\textbf{Comparison of real data and simulated data}. The solid black line represents the middle price; the red dashed line represents the relative spread from the best ask price to the middle price, and the green dashed line represents the relative spread from the best bid price to the middle price.}
  \label{drl_mm_simulation_data_curve}
\end{figure*}

We calibrate the parameters in the simulation model with the real tick data and the comparison between the simulated data and the real data from Coinbase's BTC-USD pair is demonstrated in Figure \ref{drl_mm_simulation_data_hist} and Figure \ref{drl_mm_simulation_data_curve}. As shown in Figure \ref{drl_mm_simulation_data_hist}, the real spread between the best ask price and the best bid price is central at 0.8 bps (basis points) with a standard deviation of 0.5 bps and most of the tick interval time is near 0 ms (milliseconds) and the mean is about 7 ms with a wide standard deviation of 20 ms. We can see that the simulated distribution of spread and the tick interval time is close to the real one. Moreover, the bottom two graphs show the comparison of the middle price and spread in a timeline framework. We can see that the simulated price contains trending intervals and oscillating intervals, which are the most common two scenarios in real market price analysis. 

However, due to the assumptions made in the simulation, we have lost some of the characteristics present in the actual data. For instance, when the volatility of the price increase as shown at around 00:30 of the real price graph in Figure \ref{drl_mm_simulation_data_curve}, the real bid and ask spread usually increase, since the market makers are uncertain about the current fair price and then have to quote orders conservatively. Conversely, when the price has stabilized, the real spread always becomes smaller, as demonstrated from 01:30 to 02:00 of the real price graph. But for the simulated data, the spread is always randomly sampled from the same distribution either in trending intervals or oscillating intervals as shown in the simulated price graph. 

In summary, the simulation data is able to imitate the basic characteristics of the market tick data and can further facilitate our analysis of the market making strategy in different market scenarios, which we will discuss in this section. Additionally, there are elaborate features missing in the simulation data, and for these features, we will discuss them in Section \ref{Feature_Extraction} and Section \ref{Assembled_Framework} in a real data scenario.

\subsection{Environment Design}
After finalizing the simulation, we began building the optimization pipeline for the deep reinforcement learning market-making model. To design such a model, we first need to create an environment for the agent to interact with. This environment primarily consists of the order submission and cancellation processes and the order execution process.

\subsubsection{Order Submission and Cancellation}
As with most previous research on market making, we assume that each limit order quoted by the agent is of a single unit and that the impact on price is negligible. While most of the previous work ignores the latency of the order routing process, we take into account both the latency for order submission and order cancellation. In a competitive market, latency is one of the determining factors for high-frequency trades, especially market makers, to make a profit. To be specific, low latency allows market makers to obtain more up-to-date market information, enabling them to adjust order prices or cancel orders more quickly. This ultimately reduces the risk of adverse selection. Conversely, high latency causes market makers to react very slowly to market information, resulting in their profits being eaten up by insider traders. Therefore, market makers always try their best to reduce transaction latency. While the latency stems from various aspects, such as the strategy computation module, the exchange matching engine, and the network transmission, for ease of expression we utilize an overall latency parameter $L_{submit}$ and $L_{cancel}$ to represent the sum of all kinds of latency when submitting or canceling a limit order, respectively. The specific description for submitting and canceling orders is as follows:

For the process of the order submission in our environment, from the agent submitting the order message, to the exchange receiving the order message, this period is defined as the order submission delay $L_{submit}$, and simultaneously the order status is defined as $\text{PENDING}$. Next, the exchange will determine whether the limit order is valid. The logic is if the order is a limit buy order and the order price is higher than the best ask price, or if the order is a limit sell order and the order price is lower than the best bid price, then the order is invalid, otherwise it is valid. Then, if it is valid, the order will be placed on the order book, and the order status will be $\text{OPEN}$. If it is invalid, the order status will be $\text{CLOSED}$ and the closed information will be returned to the agent.

For the process of canceling an order, from the moment the agent sends the request to cancel the order until the exchange receives the order cancellation message, this period is defined as the order cancellation delay  $L_{cancel}$, during which the order status is defined as $\text{CANCELLING}$. It should be noted that since the order still exists on the order book during the $\text{CANCELLING}$ period, it might still be executed. If the order is not executed during this period, it can be effectively canceled, otherwise the cancellation will fail.

\begin{figure*}
  \begin{center}
  \includegraphics[width=0.5\linewidth]{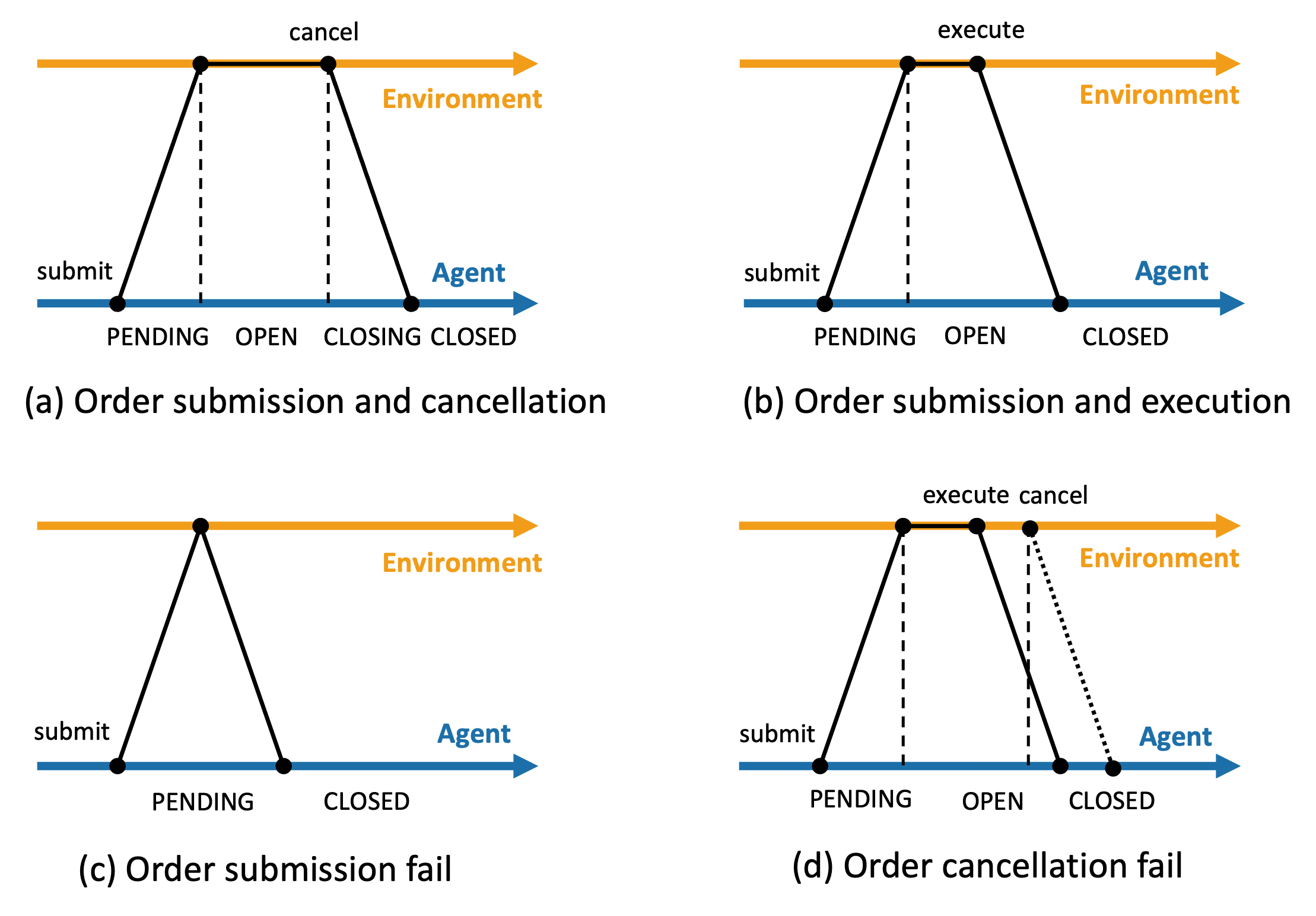}
  \end{center}
    \caption{\textbf{Diagram of limit order's life cycle}. Figure (a) shows the process of an order being submitted by an agent, received by the environment, and kept in the OPEN state, then successfully canceled by the agent; Figure (b) shows the process of an order being submitted by an agent, received by the environment, and kept in the OPEN state, then matched by the market; Figure (c) shows the process of an order being submitted by an agent but judged as a submission failure upon arrival at the environment due to failure to meet requirements; Figure (d) shows the process of an order being in the OPEN state, but failing to be canceled by the market due to delayed cancellation by the agent.}
  \label{drl_mm_simulation_order_execution}
\end{figure*}

\subsubsection{Order Execution}
Order execution modeling is typically divided into two main parts: the execution of market orders and limit orders. Market order execution modeling is relatively simple, as their execution probability is 1, and the execution price can be approximated by the price of the most recent market transaction. In contrast, while the execution price of limit orders is fixed, the execution probability is influenced by many factors, making it challenging to model limit orders accurately. As is common in literature, we simplify the logic of limit order execution modeling. The general understanding of limit order execution is that a limit order is only executed when the best opposing quote reaches or exceeds the limit order. To clarify, a limit buy order is considered executed when the best ask quote falls below the price of the simulated limit order. A limit sell order can be marked as processed once the best bid quote exceeds the price of the simulated limit order.

\subsubsection{Limit Order Life Cycle}
Having finished the limit order modeling, we now discuss the life cycle of the limit order as shown in \ref{drl_mm_simulation_order_execution}. In our environment, there are four main scenarios that may occur during the interaction between the agent and the exchange. The first scenario is when the agent sends a limit order to the exchange and then cancels the order after a period of time without it being executed, as shown in Figure \ref{drl_mm_simulation_order_execution}(a). The status of the limit order changes from $\text{PENDING}$ to $\text{OPEN}$ upon sending, and then from $\text{CANCELLING}$ to $\text{CLOSED}$ upon cancellation. The second scenario is when the agent sends a limit order to the exchange and it is executed after a period of time, as shown in Figure \ref{drl_mm_simulation_order_execution}(b). The status of the limit order changes from $\text{OPEN}$ to $\text{CLOSED}$ upon execution. The third scenario is a failed send scenario. The client sends a limit order to the exchange, but the exchange determines that the order is invalid based on the current order book price (the limit buy order's price is higher than the best ask price or the limit sell order's price is lower than the best bid price) and returns a failure message to the agent. Therefore, the order status changes directly from $\text{PENDING}$ to $\text{CLOSED}$, as shown in Figure \ref{drl_mm_simulation_order_execution}(c). The final scenario is a failed cancellation scenario where the client attempts to cancel an $\text{OPEN}$ status order on the exchange order book, but due to a delay in cancellation, the order is executed before the cancellation is completed, resulting in a failed cancellation and a change in the order status from $\text{OPEN}$ to $\text{CLOSED}$ as shown in Figure \ref{drl_mm_simulation_order_execution}(d).

\subsection{Agent Design}
After designing the logic for the environment, we proceed to design the agent logic. While the overall framework is similar to known literature, we propose a new training step interval and design an innovative state space to effectively incorporate latency factors into the agent's training process during interactions with the latency-aware environment, which enables the agent to more reasonably adjust orders to control risk.

\subsubsection{Action Frequency}
In terms of the frequency of actions on the agent side, there are currently two main types of step intervals in the literature. The first type is based on a fixed physical time interval, meaning the agent adjusts an order every few seconds or millimeters. The second type is based on a fixed number of ticks, meaning the client adjusts an order every 10 or 100 ticks. Our proposed action frequency differs from these two types. We design different step intervals based on different order statuses. Specifically, when the order status is $\text{OPEN}$ or $\text{CLOSED}$, the action frequency is one unit of tick. This means the agent can perform an action every time it receives a new tick to determine whether to adjust the current status of the order. When the order status is $\text{PENDING}$ or $\text{CANCELLING}$, the action frequency is one unit of order delay or cancellation delay. This means the agent can only perform an action when the order status changes from $\text{PENDING}$ or $\text{CANCELLING}$ to $\text{OPEN}$ or $\text{CLOSED}$, as the order status cannot be changed by the client during the $\text{PENDING}$ or $\text{CANCELLING}$ period. In short, this action frequency leads to each agent's action resulting in a change in the order status, which is more reflective of the actual situation.

\subsubsection{Action Space}
For the action space, we use a framework similar to related literature. Specifically, the action space $Action=(Action_{bid},Action_{ask})$, where $Action_{bid}$ contains 10 levels as $\{1,2…,10\}$, and $Action_{ask}$ also contains 10 levels as $\{1,2…,10\}$. Each level represents a basic price distance, and in this case, we set the basic distance to one basis point. Therefore, the agent limit order price is derived as:
\begin{equation}
\begin{split}
    p^{order\_bid} &= p^{best\_bid} \left(1-Action_{bid}\%\%\right), \\
    p^{order\_ask} &= p^{best\_ask} \left(1+Action_{ask}\%\%\right),
\end{split}
\end{equation}
where $p^{order\_bid}$ and $p^{order\_ask}$ are the agent's buy order price and sell order price, respectively. For Instance, when the level is equal to 1, it means the agent wants to quote a buy order at the distance equal to the best bid price minus one basis point ($p^{order\_bid} = p^{best\_bid} \left(1-1\%\%\right)$), and when $Action_{bid}=10$, it means the agent wants to quote a buy order at the distance equal to the best bid price minus ten basis points ($p^{order\_bid} = p^{best\_bid} \left(1-10\%\%\right)$). Conversely, the levels of $Action_{ask}$ represent the agent's desire to quote a sell order equal to the best ask price plus the corresponding number of basis points. Therefore, the action space contains a total of 100 different actions for the agent to choose from.

\subsubsection{State Space}
In this section, we describe the state space in detail. The basic state space we propose consists of four main parts. The first part is the relative distance ratio between the market's best bid and ask prices and the buy and sell limit order prices of the agent as follows: 
\begin{equation}
\begin{split}
    RDR^{bid} &= (p^{best\_bid}-p^{order\_bid})/p^{best\_bid}, \\
    RDR^{ask} &= (p^{order\_ask}-p^{best\_ask})/p^{best\_ask},
\end{split}
\end{equation}
which reflects the degree of deviation of the agent's limit order price from the market best price. For example, if the buy order is close to the best bid price and the sell order is far from the best ask price ($RDR^{bid}<RDR^{ask}$), then the buy order is more likely to be filled compared to the sell order, meaning that the agent wants to buy the asset more at this time. Conversely, if the sell order is far from the best ask price ($RDR^{bid}>RDR^{ask}$), it means that the agent wants to sell the asset. Therefore, if this state does not meet the client's expectations, the client needs to adjust the order price accordingly.

The second part is the status of the agent's order. As mentioned earlier, the agent's order status consists of $\text{PENDING}$, $\text{OPEN}$, $\text{CANCELLING}$, or $\text{CLOSED}$ and we use the one-hot encoding to represent the order status as follows:
\begin{equation}
    OS^{bid}, OS^{ask} = \left\{\begin{array}{cc}
{[1,0,0,0]} & \text{PENDING}, \\
{[0,1,0,0]} & \text{OPEN}, \\
{[0,0,1,0]} & \text{CANCELLING}, \\
{[0,0,0,1]} & \text{CLOSED},
\end{array}\right.
\end{equation}
The agent needs to adjust the sensitivity of the order based on the different states of the order. For example, when the order status is $\text{OPEN}$, the agent usually needs to be more cautious about whether the order price is reasonable because there is a risk of insider trading eating up the order. However, when the order status is $\text{CLOSED}$, the agent can patiently wait for the opportunity to open a position since there is no order in the market at this time.

The third part is the agent's inventory ratio, which reflects the size of the inventory risk that the agent is currently exposed to. The inventory ratio is defined as:
\begin{equation}
\begin{split}
    IR_{n} &= I_{n} / I_{max}, \\
    I_{n} = I_{n-1} - &Q^{exe\_ask}_{n} + Q^{exe\_bid}_{n},
\end{split}
\end{equation}
where $IR_{n}$ indicates the inventory ratio at the n-th step, $I_{n}$ represents the holding inventory of the agent at the n-th step, $Q^{exe\_ask}_{n}$ and $Q^{exe\_bid}_{n}$ are the summation of the executed ask or bid order size from the step n-1 to step n and $I_{max}$ is the maximum inventory the agent is allowed to hold.
The inventory ratio is an essential state, and the agent needs to control its inventory ratio within a relatively small range to reduce the risk of holding an inventory.

The fourth part is the agent's inventory entry price, which refers to the average price at which a trader initiates or enters a position in the asset:
\begin{equation}
    EP_{n} = \frac{I_{n-1} \times EP_{n-1} - p^{exe\_ask}_{n} \times Q^{exe\_ask}_{n} + p^{exe\_bid}_{n} \times Q^{exe\_bid}_{n}}{I_{n-1}-Q^{exe\_ask}_{n}+Q^{exe\_bid}_{n}},
\end{equation}
where $EP_{n}$ is the inventory entry price at the n-th step, $p^{exe\_ask}$ and $p^{exe\_bid}_{n}$ represents the average price for the executed order size $Q^{exe\_ask}_{n}$ and $Q^{exe\_bid}_{n}$ from step n-1 to step n, respectively.
The agent can adjust the closing order price based on the inventory entry price to control its desired profit ratio as well as the risk it is willing to take.
The above are the four basic states that we consider in the simulation stage. In the subsequent sections, we will further consider more complex states to make the reinforcement learning-based market-making strategy more profitable and stable.

\subsubsection{Reward Function}
Finally, we design the reward function as:
\begin{equation}
\begin{split}
    R_{n+1} = V_{n+1} - V_{n} - \lambda |I_{n+1}|, \\
\end{split}
\label{reward_equation}
\end{equation}
where
\begin{equation}
\begin{split}
    V_{n+1} &= I_{n+1} \times (p^{mid}_{n+1} - EP_{n+1}) + M_{n+1}, \\
    V_{n} &= I_{n} \times (p^{mid}_{n} - EP_{n}) + M_{n},\\
    M_{n+1} = M_{n} + Q^{exe\_bid}_{n+1} &(EP_{n} - p^{exe\_bid}_{n+1})+Q^{exe\_ask}_{n+1} (p^{exe\_ask}_{n+1}-EP_{n}),
\end{split}
\label{reward_explain}
\end{equation}
$R_{n+1}$ denotes the reward at step n+1, $V_{n+1}$ and $V_{n}$ are the total equity of the agent at step n+1 and n, $M_{n+1}$ and $M_{n}$ are the realized profit and loss at step n+1 and n and $\lambda$ is the hype-parameter for the penalty item of the reward.
By substituting formulas \ref{reward_explain} into equation \ref{reward_equation}, we can obtain the detailed expression of the reward function as follows:
\begin{equation}
\begin{split}
    R_{n+1} = &\underbrace{I_{n+1} \times (p^{mid}_{n+1} - EP_{n+1}) - I_{n} \times (p^{mid}_{n} - EP_{n})}_{\text{P\&L from holding inventory}} + \\
    &\underbrace{Q^{exe\_bid}_{n+1} (EP_{n} - p^{exe\_bid}_{n+1})+Q^{exe\_ask}_{n+1} (p^{exe\_ask}_{n+1}-EP_{n})}_{\text{P\&L from spread-capturing}} - \\
    &\underbrace{\lambda |I_{n+1}|}_{\text{penalty}},
\end{split}
\end{equation}
According to this formula, we can see that the agent can actually achieve profits in two ways. The first way is to obtain unrealized profits by holding inventory. For example, when the agent holds a long position and the market price rises, the agent's unrealized profits will increase, and the same is true when holding a short position and the market price falls. The second way is called spread-capturing by getting both buy and sell orders executed, thereby increasing the realized profits. As a passive market maker strategy, we encourage the second method and suppress the first method, since the first method requires the agent to hold a larger inventory, which increases the agent's inventory risk. Therefore, we add a penalty term for holding a large inventory in the reward function to suppress the holding inventory and encourage the goal of capturing the spread, also called making round-trips.

\subsection{Reinforcement Learning Algorithm}
In this work, we apply and compare different deep reinforcement learning algorithms. Deep reinforcement learning prototypes utilize a variety of implementation algorithms to tackle MDP optimization problems from different perspectives, including model-free and model-based approaches, as well as value-based and policy-based strategies, and actor-critic models that combine both policy-based and value-based learning. These models share a common feature that they leverage the powerful representation capabilities of deep neural networks (DNN) to estimate complex state-action spaces in MDP problems. Although the convergence is not guaranteed theoretically due to their complexity, there are many improvement schemes and variations proposed in the literature to ensure the algorithms' stability in practical applications,  In this section, we will introduce two representative models: the value-based model, namely Double Deep Q Network (DQN) and the actor-critic model, namely Proximal Policy Optimization (PPO).

\subsubsection{Double Deep Q Network}
Value-based models typically utilize the temporal-difference (TD) method to estimate the optimal state-action function $q_*(s, a)$, and then directly find the optimal policy $\pi^{*}$ based on the approximated state-action function. The most typical TD control algorithm is called Q-learning \cite{watkins1989learning}, and its state-action function is updated as follows:
\begin{equation}
    q\left(S_t, A_t\right) \leftarrow q\left(S_t, A_t\right)+\alpha\left[R_{t+1}+\gamma \max _a q\left(S_{t+1}, a\right)-q\left(S_t, A_t\right)\right],
\label{q_update_equ}
\end{equation}
where $q(S_{t},A_{t})$ is the value of the agent's state-action function when the agent executes action $A_{t}$ at state $S_{t}$, $R_{t+1}$ denotes the exact reward that the agent obtains after executing action $A_{t}$ at state $S_{t}$, $\max _a q\left(S_{t+1}, a\right)$ represents the future reward the agent can capture at the new state $S_{t+1}$, $\gamma$ is the discount rate and $\alpha$ denotes the learning rate. In order to optimize the state-action function, the agent needs to interact with the environment and generate enough samples for the learning iterations. 

Additionally, deep Q-networks (DQN) \cite{mnih2015human} utilize deep neural networks to approximate the state-action function. The update of the network parameters is achieved through gradients of a loss function. The loss function is  formulated based on Equation \ref{q_update_equ} as follows, and the network parameters can be updated by minimizing the loss function, accordingly.
\begin{equation}
    L_j\left(\theta_j\right)=\mathbb{E}\left[r+\gamma \max _{a^{\prime}} q\left(s^{\prime}, a^{\prime} ; \theta_{j-1}\right)-q\left(s, a ; \theta_j\right)\right]^2,
\end{equation}
where $\theta_{j}$ denotes the network parameters. Many subsequent studies have proposed various variants of the basic DQN framework for different improvement goals, such as pursuing more stable convergence, increasing exploration, and reducing variance. In this work, we focus on one of the most common variants, namely double deep Q-network (double DQN) \cite{van2016deep}. In the basic DQN model, the use of the same Q function to select and evaluate actions in the max operator of the DQN's update formula can result in an overestimation of values. To avoid this overestimation, double DQN uses two separate networks to optimize the value function: one target network for estimating the value function and one moving network for selecting actions. The update formula for the moving network is as follows:
\begin{equation}
    L_j\left(\theta_j\right)=\mathbb{E}\left[r+\gamma q^{\text{target}}\left(\underset{a}{\operatorname{argmax}}q^{\text{mov}}\left(s^{\prime}, a^{\prime} ; \theta_{j-1}\right)\right)-q^{\text{mov}}\left(s, a ; \theta_j\right)\right]^2,
\end{equation}
Moreover, the parameters of the target network are updated by a periodic copy of the weights of the moving network. By decoupling the max operator, the double DQN algorithm shows a reduction in the observed overestimation.

\subsubsection{Proximal Policy Optimization}
The Actor-Critic (AC) method combines policy-based and value-based methods. This method employs two separate networks to represent the policy and value functions. The AC method consists of two modules: the critic and the actor. The critic updates the parameters of the value network, while the actor updates the parameters of the policy network based on the value function provided by the critic network. By introducing a baseline, namely the advantage function, on top of the policy-based method, the AC method reduces the variance of the policy-based method. Therefore, the general update formula for the AC method is as follows:
\begin{equation}
    \theta \leftarrow \theta+\alpha \nabla_\theta \log \pi_\theta(s, a) q_w(s, a),
\end{equation}
where $\theta$ is the actor network's parameters, $w$ is the critic network's parameters. A variety of choices of actor and critic networks are proposed in literature \cite{schulman2017proximal}. 

The Actor-Critic (AC) method introduces two different networks that must interact and update with each other. However, this interaction can cause the estimation error to gradually increase, which can lead to instability in the training process. To address this issue, the Proximal Policy Optimization (PPO) algorithm proposed in \cite{schulman2017proximal} designs a new objective function that enables multiple epochs of minibatch updates. The objective function is shown below:
\begin{equation}
    L^{\text{CLIP}}(\theta)=\hat{\mathbb{E}}_t\left[\min \left(r_t(\theta) \hat{A}_t, \operatorname{clip}\left(r_t(\theta), 1-\epsilon, 1+\epsilon\right) \hat{A}_t\right)\right]
\end{equation}
where $r_t(\theta)=\frac{\pi_\theta\left(a_t \mid s_t\right)}{\pi_{\theta_{\text {old }}}\left(a_t \mid s_t\right)}$ denotes the probability ratio of new trajectory over the old trajectory. According to \cite{schulman2017proximal}, the PPO algorithm has the benefits of trust region policy optimization (TRPO) \cite{schulman2015trust} but is much simpler and more general.

\begin{figure*}
  \begin{center}
  \includegraphics[width=0.6\linewidth]{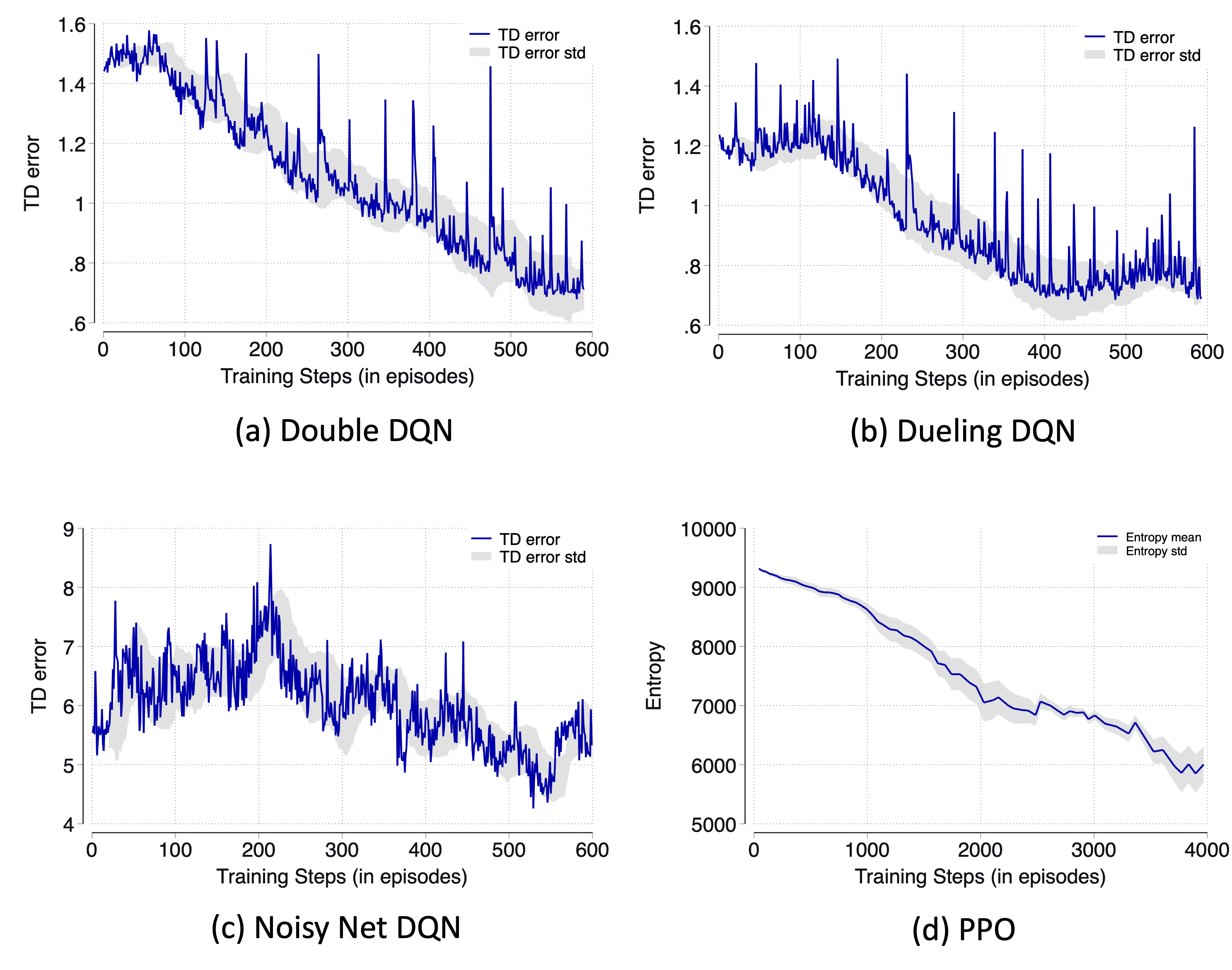}
  \end{center}
    \caption{\textbf{Training phase loss for different algorithms}. Figures (a), (b), and (c) show the changes in TD error during the training process of market-making strategies based on different types of DQN. In these value-based models, TD error can indirectly reflect the convergence of the model. Figure (d) shows the changes in the output action entropy during the training process of the market-making model based on the PPO algorithm. In these actor-critic-based models, entropy can be used to represent its convergence.}
  \label{drl_mm_simulation_train_loss}
\end{figure*}

\subsection{Simulation Results}
In this chapter, we present the results of our simulation experiments, which aimed to evaluate the performance of the proposed deep reinforcement learning model for high-frequency market making strategies. The framework includes a tick-level environment and a latency-aware agent and trained the model using the DQN and PPO algorithms. In addition, we compared the performance of our model with two baseline algorithms: a fix-spread algorithm and a simple version of the Avellaneda-Stoikov algorithm. Moreover, we also evaluate the performance of the proposed model under different environment settings.

\begin{figure*}
  \begin{center}
  \includegraphics[width=.6\linewidth]{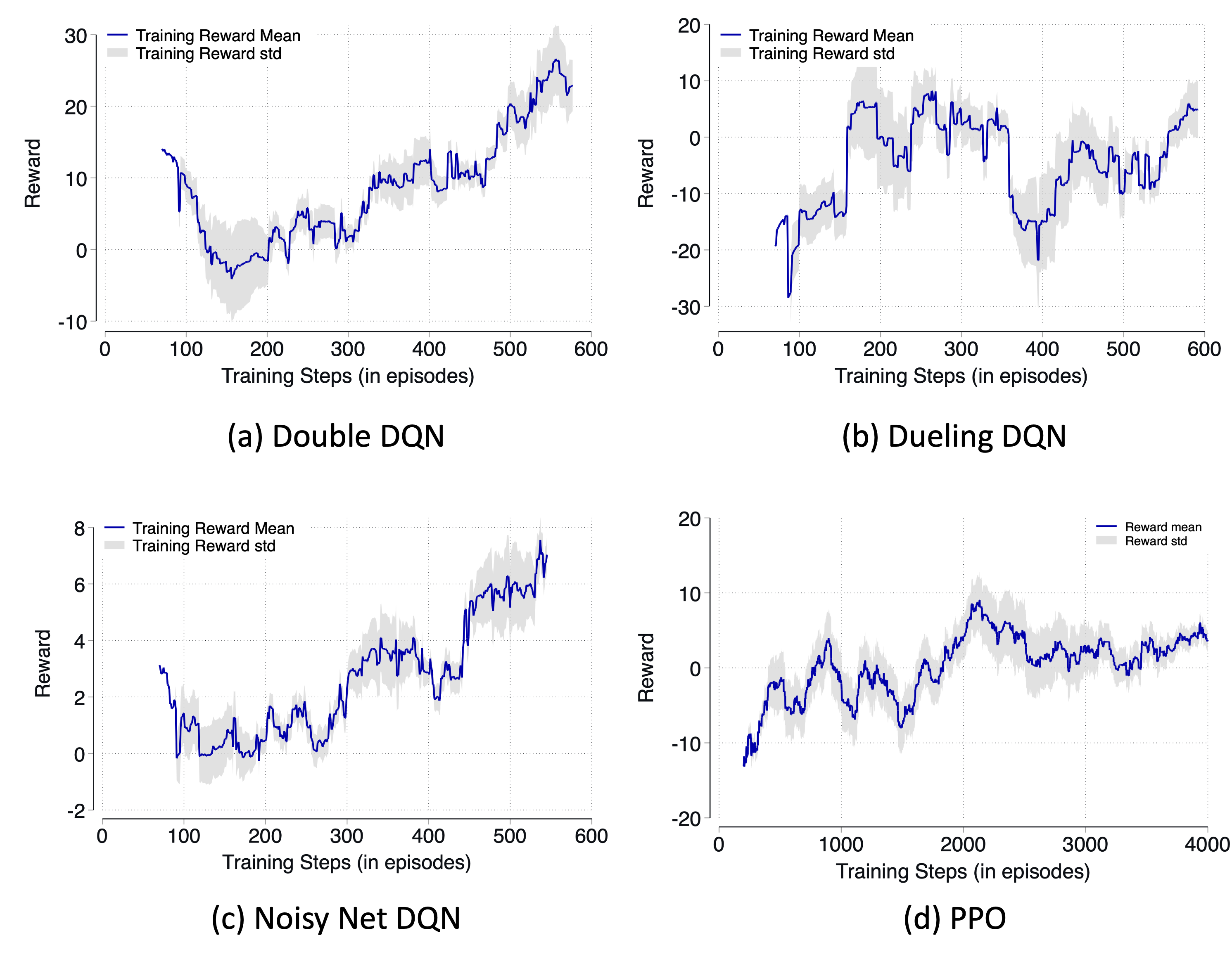}
  \end{center}
    \caption{\textbf{Training phase reward for different algorithms}. The figure shows the cumulative profit for market-making strategies based on different DRL models over each episode during the training process. This profit does not include the penalty term in the loss function.}
  \label{drl_mm_simulation_train_reward}
\end{figure*}

\subsubsection{Experimental Setup}
The simulation model is calibrated with the historical data collected from the Coinbase exchange for two months, from September to October 2022. The tick-level environment was designed based on the simulated top order book data, and the latency-aware agent was trained based on different optimization algorithms to quote bid and ask orders based on the current state of the order book and the agent conditions. We used a variety of hyperparameters for each algorithm and evaluated the performance of the model using several metrics, including average profit per trade, number of successful trades, and average bid-ask spread.

For the training phase, we trained the model for a fixed number of episodes, like 600 episodes. At the beginning of each episode, the initial inventory is zero. Then, with the execution of the agent's bid and ask orders during the interaction with the environment, the inventory accumulation begins. Once the inventory returns to zero, it means this episode has ended. Therefore, the steps or iterations in each episodes are different based on the duration time of non-zero inventory. For the testing phase, in order to fairly compare different algorithms, we simulate a set of test data with a fixed time length, such as 8 hours, and then compare the performance of different algorithms based on the overall results in the fixed test dataset. For the hyperparameters of the deep reinforcement learning models, we use a learning rate of 0.00005 for the DQN algorithm and 0.00010 for the PPO algorithm and a discount factor of 0.99 for both algorithms. Additional, we also use a replay buffer of capacity 1000000 iterations for the DQN algorithm and a batch size of 8192 for both  algorithms.

For the fix-spread algorithm, we set the bid and ask prices at a fixed spread around the mid-price of the order book all the time and we choose the median spread in the action space as the fix spread. For the simplified Avellaneda-Stoikov algorithm, we assume its volatility parameter is a constant, so in this model, the order price is only related to the inventory size. We further simplify the order price and inventory size as a linear proportional relationship, so we only need to determine its proportional parameter, where we use enumeration methods to select the specific values of the parameter during the training phase.

\begin{figure*}
  \begin{center}
  \includegraphics[width=.6\linewidth]{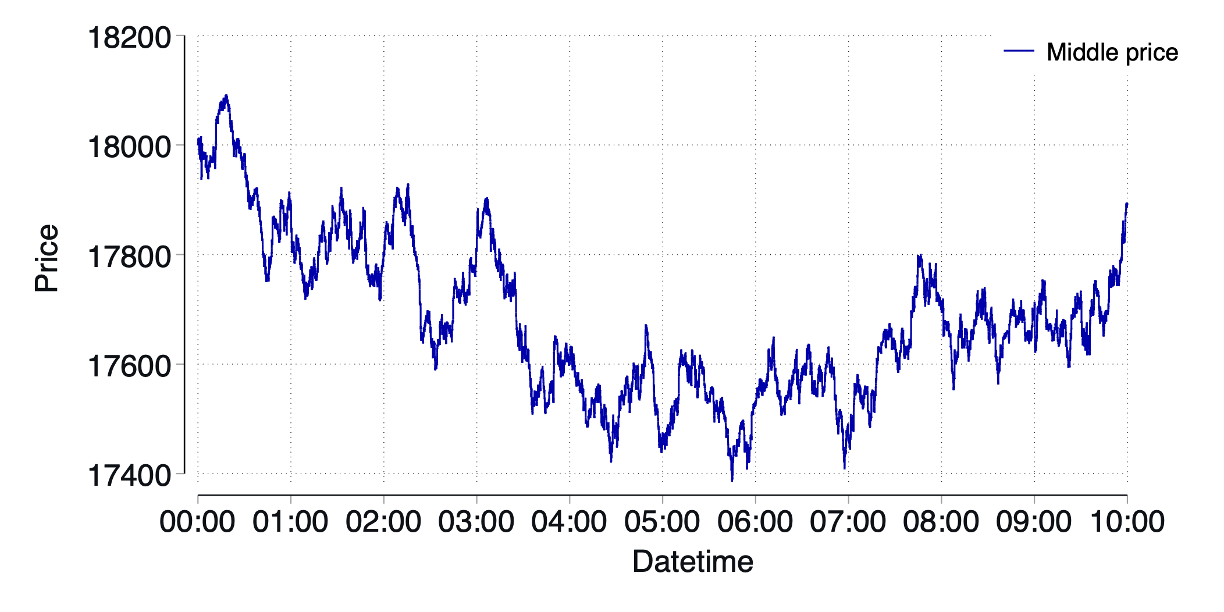}
  \end{center}
    \caption{\textbf{The simulated testing tick level data}. The simulated testing dataset contains tick data for a continuous period of 10 hours.}
  \label{drl_mm_simulation_test_price}
\end{figure*}

\subsubsection{Results}

\textbf{Training Phase:} To compare the convergence of different reinforcement learning algorithms, we observed training indicators during the training process. For the DQN-based models, we used TD error to determine whether it converged, while for the PPO model, we judged its convergence by observing the entropy of the probability of its action space. As shown in Figure \ref{drl_mm_simulation_train_loss}(a)(b)(c), the TD error of the DQN-based model tended to stabilize around 600 episodes, with the TD error of Double DQN and Dueling DQN stabilizing around 0.8. While as for the noisy net DQN, due to the noisy terms in its network structure, its TD error was relatively large, as shown in the graph at episode 600, where its TD error was 5.5. For the PPO model, as shown in Figure \ref{drl_mm_simulation_train_loss}(d), it can be seen that more training steps are required to achieve stability. This is because PPO is updated based on the on-policy mode, which means that each time the model is updated, the data used is a trajectory obtained under the latest policy parameters. When the policy parameters are updated, new trajectories need to be generated as training data. It can be observed that the entropy of the PPO model gradually stabilizes at around 4000 episodes.

\begin{figure*}[t]
  \begin{center}
  \includegraphics[width=.6\linewidth]{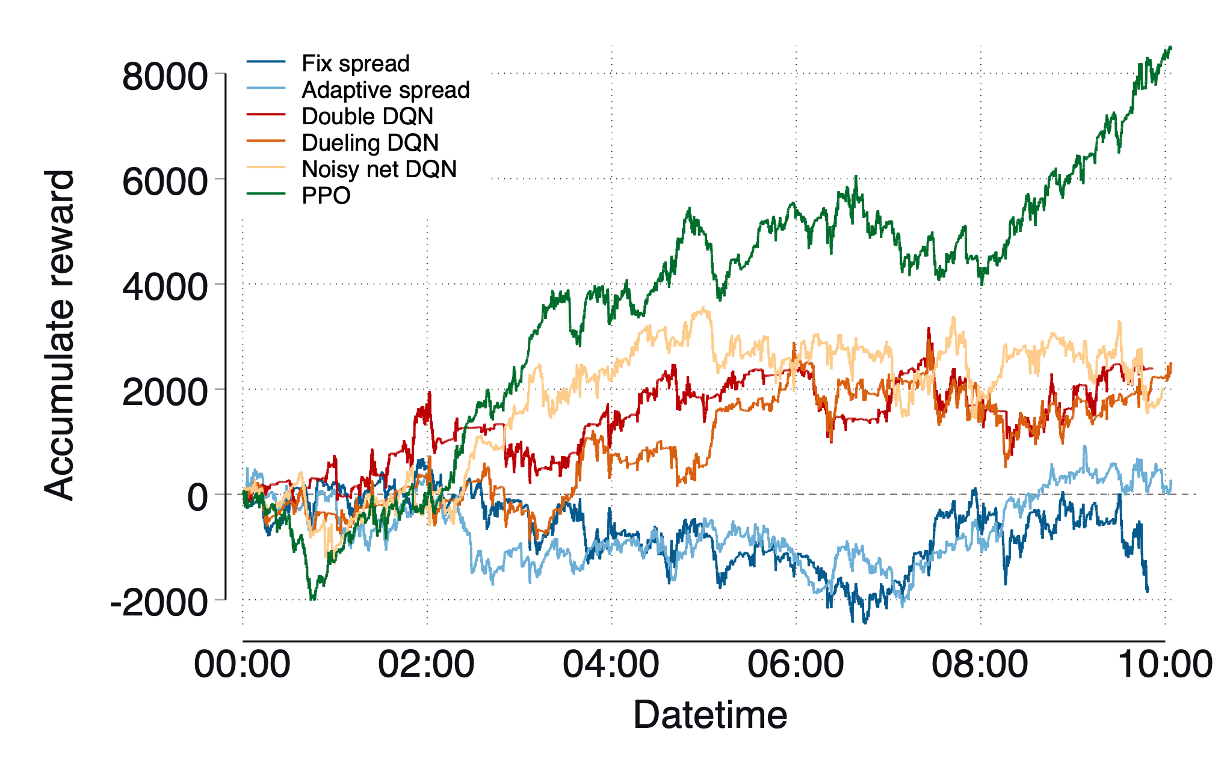}
  \end{center}
    \caption{\textbf{Performance of different market making strategies under the simulated testing data}. The figure shows the cumulative profit of different strategies on testing data. The cumulative profit of the Fix Spread and Adaptive Spread strategies are low, while the strategy based on DQN has increased cumulative returns compared to the first two strategies. Finally, the best-performing strategy is based on the PPO algorithm.}
  \label{drl_mm_simulation_test_cumreward}
\end{figure*}

Meanwhile, we also recorded the changes in reward for each episode during the training process. As shown in Figure \ref{drl_mm_simulation_train_reward}, the reward of the DQN-based model gradually increased with the increase of training steps. Compared to the other two DQN models, the reward of Double DQN was more stable and reached a final value of 22, which is relatively high. For the PPO model, the reward of each episode increases slowly to about 8 with the increase of training steps, and its standard deviation also becomes noticeably smaller.

\textbf{Test Phase:} During the testing phase, we simulated tick-level test data with calibration parameters consistent with the training data, as shown in Figure 1. The test data is approximately 10 hours long and consisted of a total of 240,000 tick data, with each tick including the best bid price and best ask price, as well as the current timestamp. The middle price in the figure represents the arithmetic average of the best bid price and best ask price.

Under this test data, we compared various high-frequency market-making models based on different deep reinforcement learning algorithms and baseline algorithms. Firstly, and most importantly, we compared the cumulative reward of various models on this test data, as shown in Figure \ref{drl_mm_simulation_test_cumreward}. From the figure, we can see that models based on deep reinforcement learning generally have higher cumulative rewards than baseline models. Among them, the model based on the PPO algorithm can achieve the highest cumulative reward, followed by three models based on DQN algorithms. The cumulative reward of the baseline based on fix spread has large fluctuations, while the baseline based on adaptive spread has reduced fluctuations but also reduced actual profits. From the timeline perspective, within the first two hours of testing data, reinforcement learning-based market making models cannot achieve good rewards. However, in the next three hours, models based on PPO and Noisy Net DQN can quickly accumulate relatively high rewards, while models based on PPO and Dueling DQN perform better in the last two hours of testing data. Overall, models based on PPO have significant improvements in both cumulative rewards and volatility compared to other models.

\begin{figure*}
  \begin{center}
  \includegraphics[width=.6\linewidth]{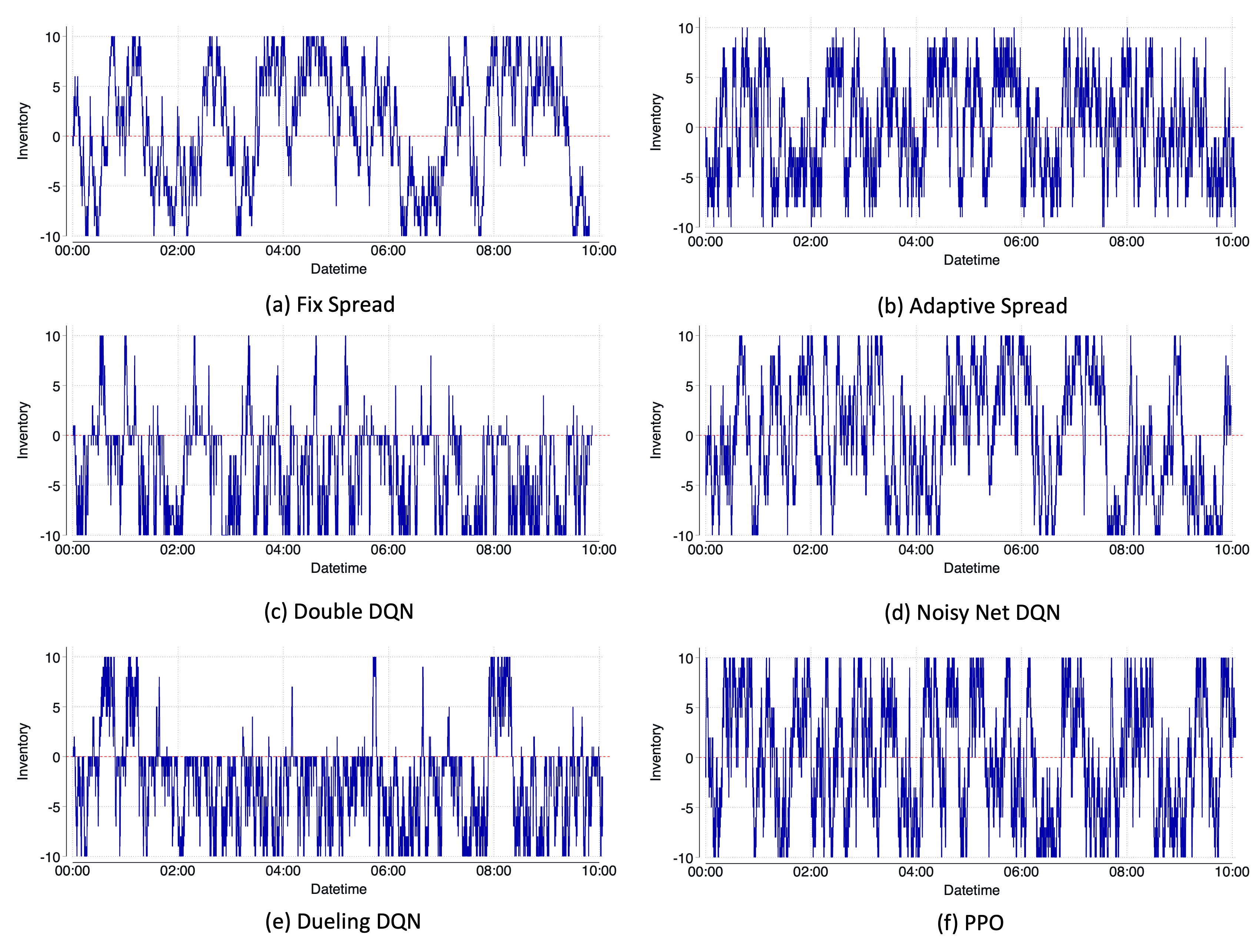}
  \end{center}
    \caption{\textbf{Inventory trajectory during the execution of market making strategies in the testing data}. Figure (a) shows that the Fix Spread strategy cannot effectively control inventory near 0, while other types of strategies can. Figures (c) and (e) show that the Double DQN and Dueling DQN strategies tend to hold negative inventory to survive the market decline. Figure (f) shows that the change amplitude and frequency of inventory in the PPO model are relatively large compared to other strategies.}
  \label{drl_mm_simulation_test_position}
\end{figure*}

Meanwhile, we also observed the changes in inventory for each model, as shown in Figure \ref{drl_mm_simulation_test_position}. It's obvious that the model based on fixed spread did not control inventory. Therefore, while the market was in a volatile phase, this model could generate profits. However, when the market was in a trending phase, this model accumulated a relatively large inventory for a long time, thus increasing inventory risk. On the contrary, the model based on adaptive spread could control the size of inventory at any time, but reduced the risk, resulting in lower profits or even losses. As for the model based on deep reinforcement learning, we can see that this type of model can basically pursue positive rewards while controlling inventory.

\begin{figure*}
  \begin{center}
  \includegraphics[width=0.5\linewidth]{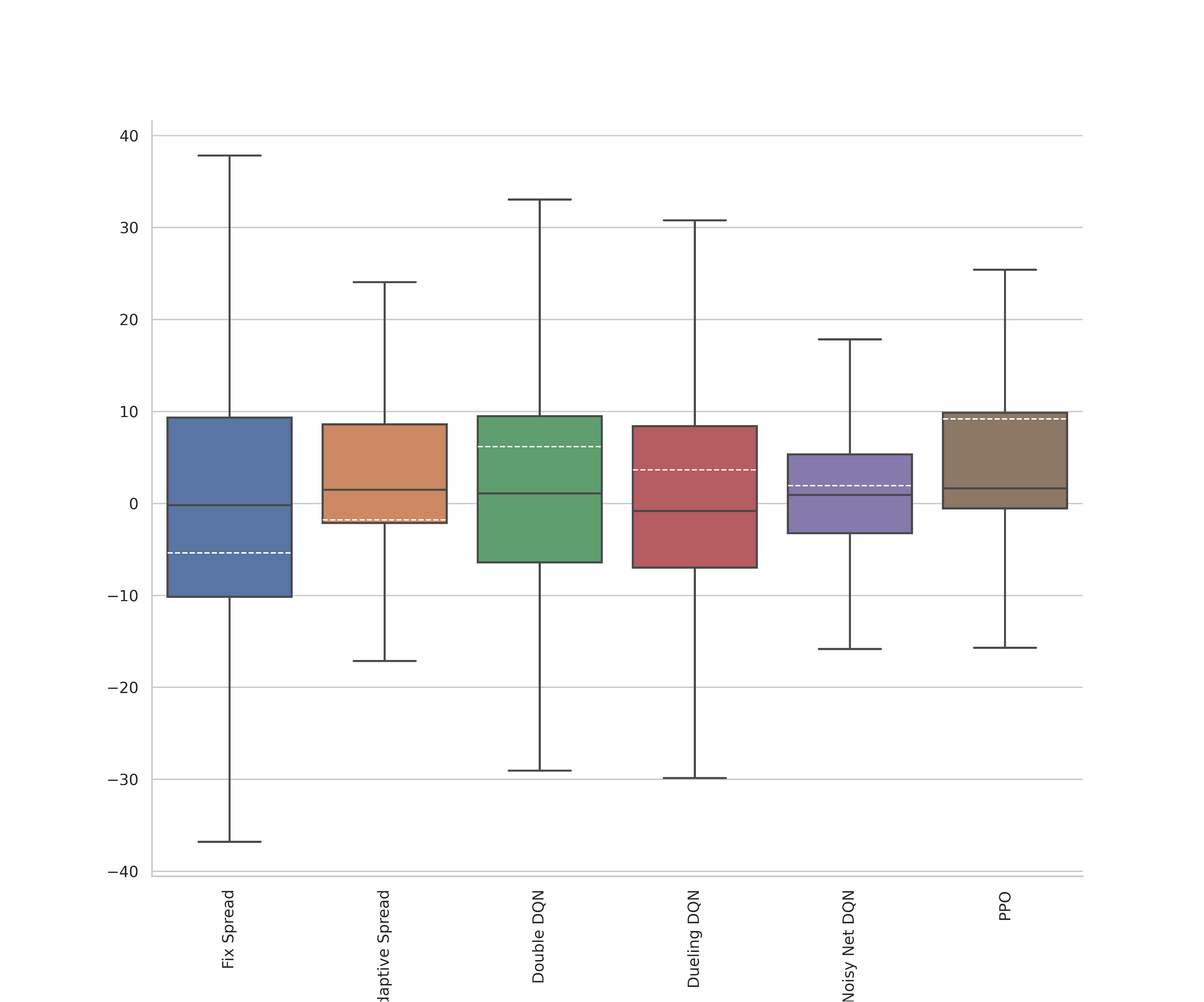}
  \end{center}
    \caption{\textbf{Boxplot of reward in each episode for different market making strategies}. The figure shows the percentile and mean of rewards for each episode under different models, where the white dashed line represents the mean. It can be seen that the reward fluctuation amplitude of Fix Spread is the largest, and the mean is also the lowest; policies based on DQN can achieve positive average rewards; policies based on PPO can achieve the highest average rewards, and the change in rewards is relatively small.}
  \label{drl_mm_simulation_test_game_reward}
\end{figure*}

\begin{figure*}
  \begin{center}
  \includegraphics[width=0.5\linewidth]{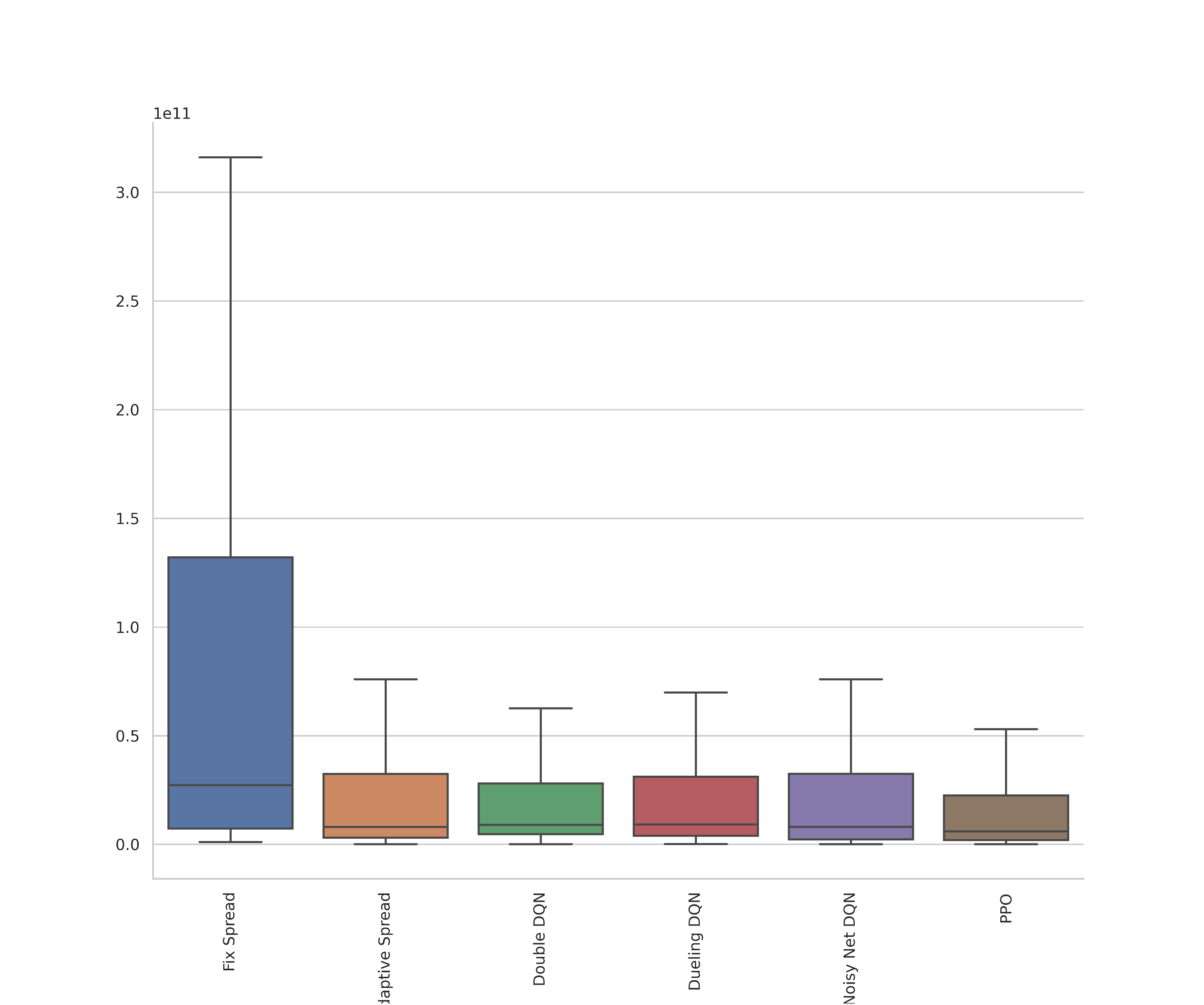}
  \end{center}
    \caption{\textbf{Boxplot of duration in each episode for different market making strategies}. The figure shows the duration of each episode under different models. It can be seen that the Fix Spread model has the longest duration because inventory is not controlled, while the duration of other models is relatively shorter because they have the inventory control module.}
  \label{drl_mm_simulation_test_game_duration}
\end{figure*}

Additionally, we compare and analyze the performance of each market-making model on an episode basis from a statistical perspective. As shown in Figure \ref{drl_mm_simulation_test_game_reward} and Figure \ref{drl_mm_simulation_test_game_duration}, we found that under the fix spread model, the overall reward fluctuates the most in each episode, and its holding time is the longest. This is because it does not have measures to control inventory, resulting in a large and long-lasting inventory. Therefore, most of its rewards come from holding inventory, rather than from buying and selling price differences. As a result, its income fluctuates significantly with price fluctuations. Under the adaptive spread model, the overall reward of each episode has decreased, and the holding time has also decreased. However, its average reward is still negative because adaptive spread does not take into account the actual reward. Further, in the three DQN models, the average reward is positive, and the holding time is maintained at a relatively low level, indicating that the strategy is actually profitable. Among them, the model based on noisy net DQN has relatively smaller reward fluctuations, which means that the risk is reduced, but its profit is relatively smaller than other DQN models. Finally, the PPO model can control the volatility of rewards to achieve relatively high average returns. This indicates that the strategy's income under the PPO model mainly comes from buying and selling price differences. Therefore, its holding time is the shortest, and the proportion of income obtained by holding inventory is relatively small. Instead, the proportion of income obtained through buying and selling price differences is relatively high, and the income from buying and selling price differences is more stable than that from inventory, and it does not fluctuate significantly with price fluctuations.

\begin{figure*}
  \begin{center}
  \includegraphics[width=0.5\linewidth]{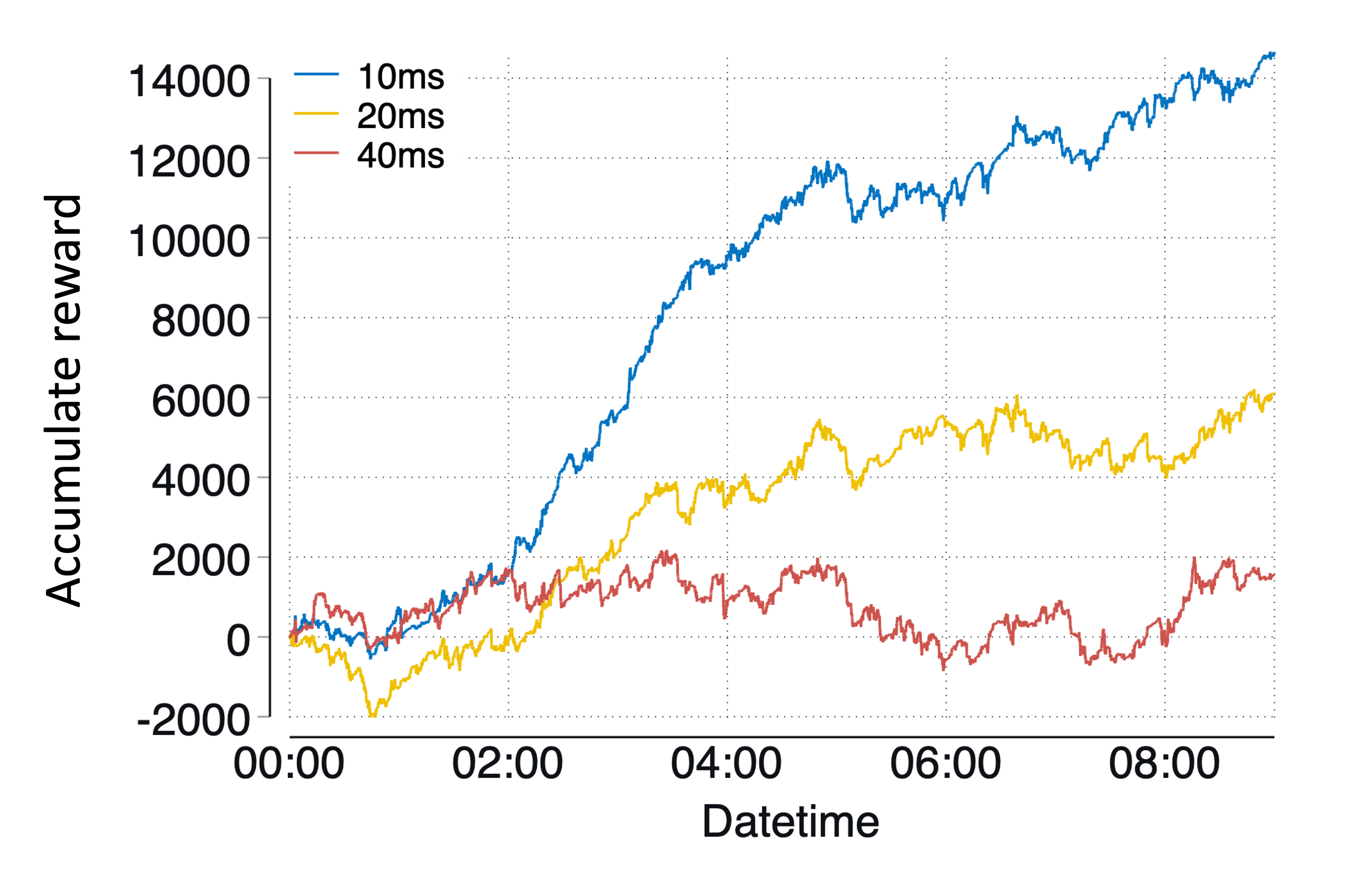}
  \end{center}
    \caption{\textbf{Accumulative reward of the PPO model under different latency for order placement}. From the chart, it can be seen that as the delay in placing orders increases, the cumulative returns of the strategy gradually decrease.}
  \label{drl_mm_simulation_test_latency_create_order_pnl}
\end{figure*}

\begin{figure*}
  \begin{center}
  \includegraphics[width=0.5\linewidth]{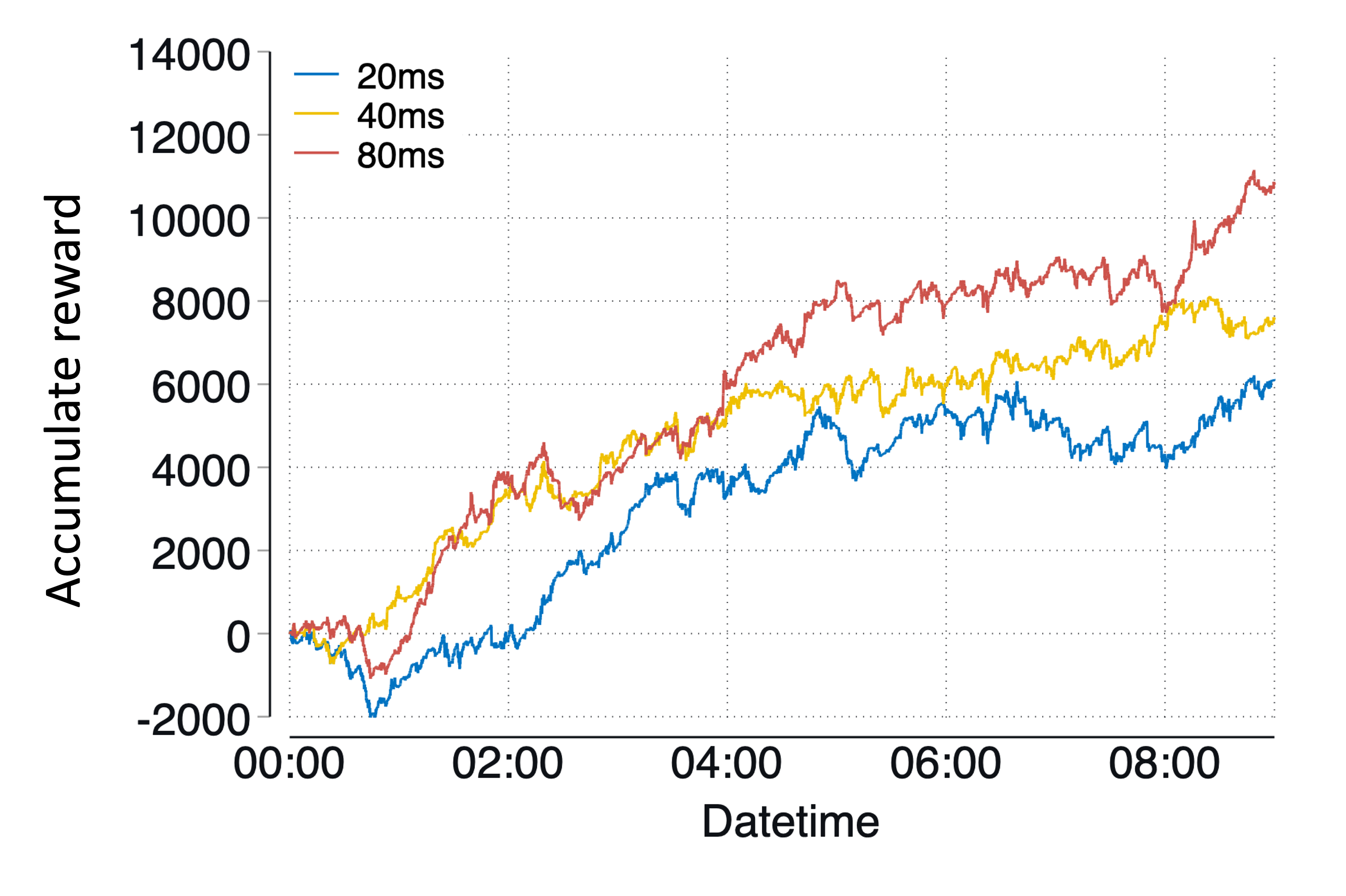}
  \end{center}
    \caption{\textbf{Accumulative reward of the PPO model under different latency for order cancellation}. This graph shows that as the cancellation delay increases, the cumulative profit of the strategy actually increases, which is opposite to the phenomenon of order placement delay.}
  \label{drl_mm_simulation_test_latency_cancel_order_pnl}
\end{figure*}

\textbf{Order Placement Latency Analysis}: 
In latency analysis, according to the tick-level model we have established, we can separate latency into order placement latency and order cancellation latency for separate analysis. Considering that the PPO model performs the best in simulation experiments, we use it as an example to analyze latency. As shown in Figure \ref{drl_mm_simulation_test_latency_create_order_pnl}, it can be seen that there is a significant difference in the performance of market-making strategies under different order placement latency, and as the order placement latency decreases, the consistency of the strategy's performance improves. When the order placement latency is reduced to 10ms, the strategy's cumulative profit reaches 15000, while when the order placement latency increases to 40ms, the strategy's cumulative profit decays to around 1000. This indicates that order placement latency has a significant impact on the profit of high-frequency market making strategies, for various reasons such as missing short-term trading opportunities in the market if the order placement latency is too high, ultimately leading to a decrease in profit.

\begin{figure*}
  \begin{center}
  \includegraphics[width=0.5\linewidth]{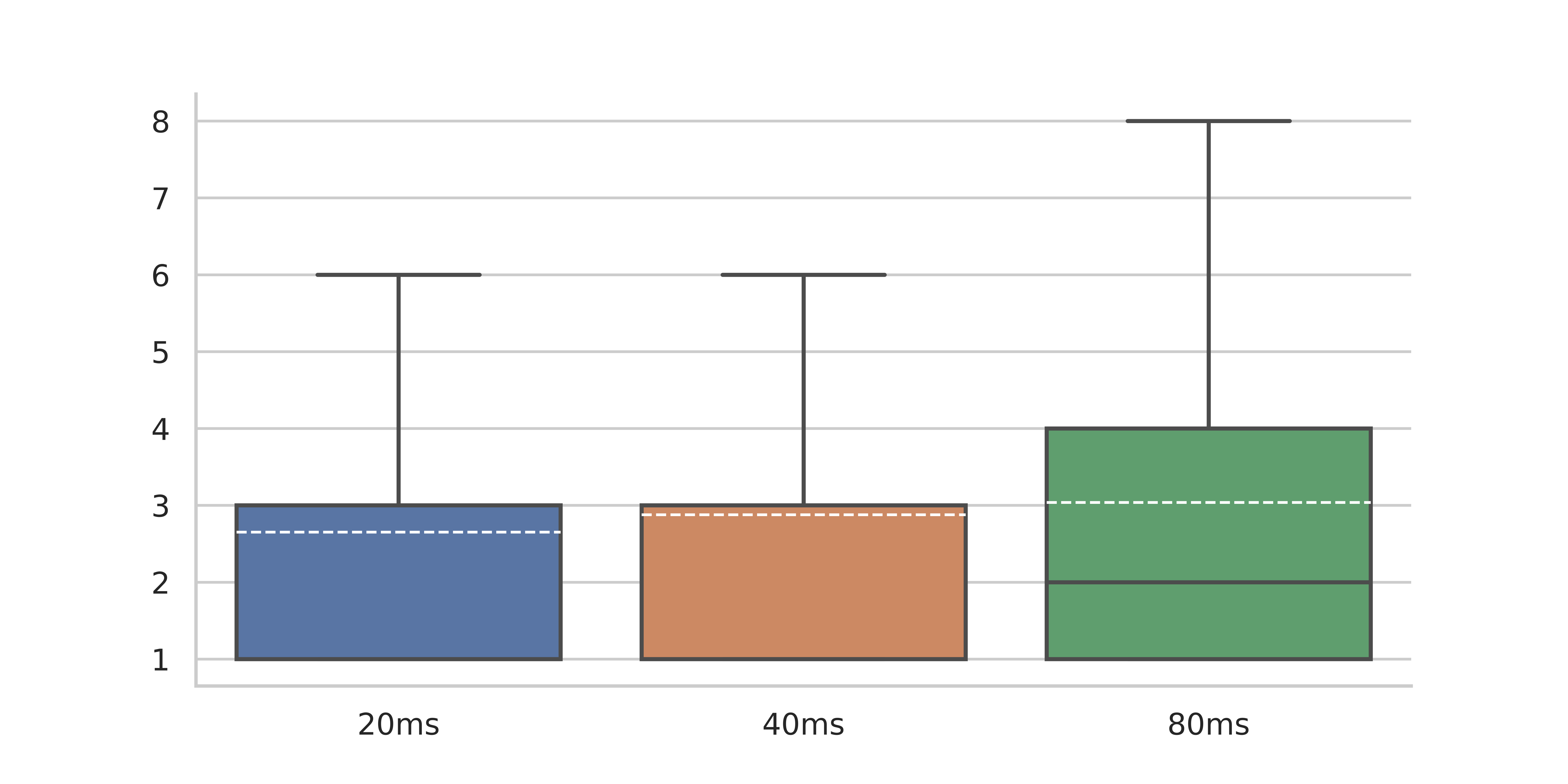}
  \end{center}
    \caption{\textbf{Boxplot of maximum Inventory in each episode under different order cancellation latency}. From the box plot, it can be seen that as the cancellation delay increases, the mean of the maximum inventory of the strategy in each episode (white dashed line) gradually increases, indicating that the strategy is taking on higher risks and therefore able to obtain greater rewards.}
  \label{drl_mm_simulation_test_latency_cancel_order_maxposition_box}
\end{figure*}

\textbf{Order cancellation Latency Analysis}: 
When analyzing the delay of order cancellation, we observe different situations. As shown in Figure \ref{drl_mm_simulation_test_latency_cancel_order_pnl}, when the cancellation delay increases, the profit of the market-making strategy gradually increases. Specifically, when the delay increases from 20ms to 80ms, the strategy's cumulative profit increases from 8000 to 14000, which is exactly the opposite of the results of the order placement delay analysis. For this phenomenon, we can explain it to some extent from Figure \ref{drl_mm_simulation_test_latency_cancel_order_maxposition_box}. Figure \ref{drl_mm_simulation_test_latency_cancel_order_maxposition_box} shows the statistical data of the maximum inventory achieved by the strategy in each episode. It can be seen that when the cancellation delay increases, the strategy tends to hold a larger inventory. This is because as the cancellation delay increases, the strategy's closing operation becomes less sensitive. Specifically, when the strategy holds a positive inventory and the price continues to fall, the strategy should cancel the current sell order and lower its price. However, due to the large cancellation delay, the price rebounds and rises when the order is being cancelled, so the actual price at which the sell order is executed is higher than the price at which the sell order will be lowered. Therefore, it increases the profit of the strategy.

\subsubsection{Discussion}
Our simulation experiments demonstrate the potential of deep reinforcement learning for high-frequency market making strategies and provide insights into the factors that influence the performance of such models. We believe that our findings could have important implications for both academic research and practical applications in the cryptocurrency and financial industries.

One of the strengths of our model is its ability to quickly adapt to changing market conditions. We found that the DQN algorithm was particularly effective in this regard, as it was able to update its policy based on recent experiences. However, the PPO algorithm tended to outperform the DQN algorithm in terms of overall profitability.

One limitation of our study is that our simulations were based on historical data, and may not accurately reflect future market conditions. In addition, our model was trained and tested using a single cryptocurrency exchange, and may not generalize to other exchanges or markets. Finally, our evaluation metrics focused primarily on short-term profitability and did not account for longer-term risks or market trends.

\section{Feature Extraction}\label{Feature_Extraction}
After completing the basic reinforcement learning framework and verifying its effectiveness on simulation data, we will now verify the performance of this model on real data. Unlike simulation data, many signals or features with price prediction capabilities, also known as alpha signals, can be discovered from real data. Especially for high-frequency trading, these predictive signals extracted from real data can help high-frequency traders accurately predict market behaviors and dynamics in a timely manner, thus maximizing their profits and controlling risks. Specifically for market makers, when alpha signals predict that the price may rise, the limit sell order price can be adjusted upward to reduce the risk of adverse selection. This chapter focuses on extracting these signals or features from historical real data, to help further improve the performance of the reinforcement learning strategy in subsequent chapters. These features include order book imbalances, trade imbalances, and past returns. We will discuss each feature in detail, including its inherent logic and its measure of future price prediction ability.

\subsection{Order Book Imbalances}
Many existing literature has proven that the shape of the order book has a significant predictive effect on future price returns \cite{bouchaud2002statistical}\cite{alfonsi2010optimal}\cite{cont2021price}. The shape of the order book is determined by the prices and quantities of limit buy and sell orders. When the shape of the order book deviates from the equilibrium state, it means the balance between buyers and sellers in the market is disrupted, which may lead to corresponding price changes. For example, when the average distance between buy order prices and the best bid price is farther than that of sell order prices, it means that sellers in the market are more aggressive than buyers, therefore the probability of prices falling in the short term is greater. Similarly, when the average quantity of buy orders is greater than that of sell orders, it means that more people in the market want to buy, therefore the probability of prices rising in the short term is greater. However, there are many different methods for defining average distance, average quantity, and order book imbalance. In this work, we apply the following formula to evaluate the order average price distance imbalance:
\begin{equation}
\begin{split}
    &\mathrm{OIDA}_t^{i} =p^{a}_{t}\left(Q_i\right)-p^{a}_{t}(1),\\ 
    &\mathrm{OIDB}_t^{i} =p^{b}_{t}(1)-p^{b}_{t}\left(Q_i\right),\\ &\mathrm{OID}_t=\sum_i^{N}\left(\frac{\mathrm{OIDA}_t^{i}}{\mathrm{OIDB}_t^{i}}-1\right) / N,
\end{split}
\end{equation}
where $Q_i$ denotes the i-th element in an accumulated quantity array, say $Q_i\in\{10,20,40,...\}$, $p^{a}_{t}\left(Q_i\right)$ and $p^{b}_{t}\left(Q_i\right)$ is the ask or bid price at the level that the accumulated quantity from the best ask or bid level is equal to $S_i$, $\mathrm{OIDA}_t^{i}$, $\mathrm{OIDB}_t^{i}$ is the i-th distance of the ask side and the bid side, $\mathrm{OID}_t$ is the overall order price distance imbalance ratio. From the above formula, it can be seen that we have integrated multi-level order price distance imbalance. Therefore, compared to only considering the price imbalance of a single level, this indicator can reflect the overall shape of the order book and is more robust.

\begin{figure*}
  \begin{center}
  \includegraphics[width=.5\linewidth]{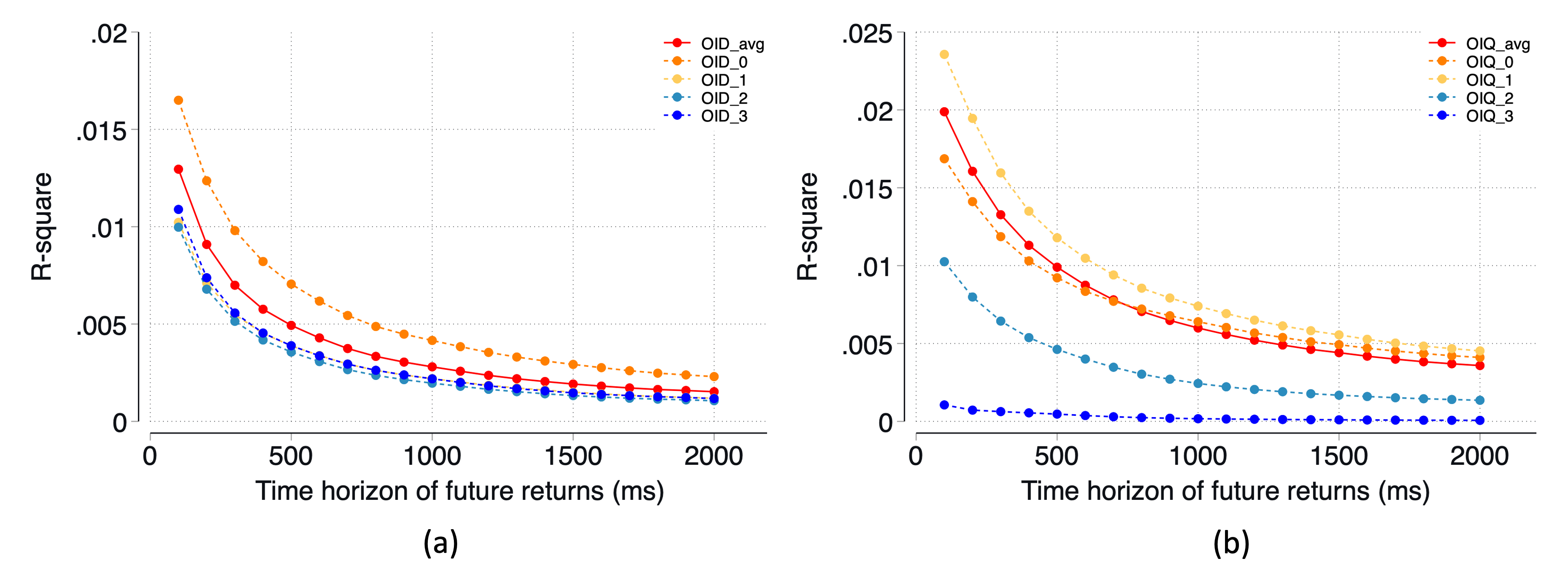}
  \end{center}
    \caption{\textbf{Comparison of $R^2$ values of order book imbalance indicators at different time horizons.} (a) Order book imbalance of distance (OID): The solid red line represents the average OID, while the other dashed lines represent the OID at different $Q_i$ values. (b) Order book imbalance of quantity (OIQ): Similarly, the solid red line is the average OIQ and the other dashed lines represent the OIQ at different $D_i$ values.}
  \label{drl_mm_oi_r_square}
\end{figure*}

Additionally, we define an alternative order book imbalance indicator, namely the order average quantity imbalance as follows:
\begin{equation}
\begin{split}
    &\mathrm{OIQA}_t^{i} =q^{a}_{t}\left(D_i\right),\\ 
    &\mathrm{OIQB}_t^{i} =q^{b}_{t}\left(D_i\right),\\ &\mathrm{OIQ}_t=\sum_i^{N}\left(\frac{\mathrm{OIQA}_t^{i}}{\mathrm{OIQB}_t^{i}}-1\right) / N,
\end{split}
\end{equation}
where $D_i$ denotes the i-th price distance ratio in an order price array, say $D_i\in\{5\%\%,10\%\%,20\%\%,...\}$, $q^{a}_{t}\left(D_i\right)$ and $q^{b}_{t}\left(D_i\right)$ is the ask or bid accumulated quantity from the best level to the level that the order price distance ratio is equal to $D_i$, $\mathrm{OIQA}_t^{i}$, $\mathrm{OIQB}_t^{i}$ is the i-th accumulated quantity of the ask side and the bid side, $\mathrm{OIQ}_t$ is the overall order quantity imbalance ratio. Similar to the order price distance imbalance, this indicator can also reflect the overall shape of the order book. When the cumulative quantity of the sellers is greater than the buyers, i.e. the indicator is positive, the predicting probability of a price drop is higher. Therefore, this indicator is also one of the characteristics of order book imbalance.

To validate the predictive ability of the two indicators of order book imbalance, we compared their explanatory power for future returns. Specifically, we defined the following univariate linear functions:
\begin{equation}
\begin{split}
    \text{fret}_t=\alpha+\beta\mathrm{OID}_t+\epsilon_t, \\
    \text{fret}_t=\alpha^{\prime}+\beta^{\prime}\mathrm{OIQ}_t+\epsilon^{\prime}_t,
\end{split}
\end{equation}
where $\text{fret}_t$ denotes the future returns at different time horizons, say 100ms, 200ms, 500ms, 1000ms, 2000ms. We use OLS to fit the above linear functions and compare $R^2$ values at different time horizons, as shown in Figure \ref{drl_mm_oi_r_square}. From Figure \ref{drl_mm_oi_r_square}(a), we can see that as $i$ increases and $Q_i$ values become larger, $R^2$ of the indicator OID consistently decreases for future returns with a longer time horizon, indicating a decreasing predictive ability over time horizons. And the best performance is achieved at $Q_1$. However, we use the average OID in practice since it is more robust. The same declining trend can also be seen in Figure \ref{drl_mm_oi_r_square}(b), which means that the predictive ability of the second indicator OIQ also decreases over time. Moreover, compared to OID, OIQ can obtain a larger $R^2$ value, which infers that the predictive ability of OIQ is relatively stronger than that of OID.

\subsection{Trade Flow Imbalances}
In addition to order book information, exchange-provided trade flow information can also generate many useful predictive indicators \cite{cont2021price}\cite{cont2014price}. Trade flow refers to the information formed after the market orders and the limit orders on the order book are matched, which can provide the most timely and latest transaction direction, price, and volume. Based on trade flow information, we can construct various predictive indicators. In this chapter, we mainly discuss the following three types. 

The first type is the volume imbalance rate, which predicts future returns by calculating the difference between the active buy trade volume and the active sell trade volume in the past period. The logic is that if the active buy trade volume is higher than the active sell trade volume, it means that the buy side traders temporarily dominate, so the probability of predicting future price increases is relatively high. The definition of the volume imbalance rate is as follows:
\begin{equation}
    \mathrm{TIQ}_t^{\delta}=\frac{\mathrm{TIQS}_{[t-\delta, t]}-\mathrm{TIQB}_{[t-\delta, t]}}{\mathrm{TIQS}_{[t-\delta, t]}+\mathrm{TIQB}_{[t-\delta, t]}},
\end{equation}
where $\delta\in\left\{100ms, 200ms, 500ms, 1000ms, 2000ms\right\}$ denotes the past time horizons, $\mathrm{TIQS}_{[t-\delta, t]}$ and $\mathrm{TIQB}_{[t-\delta, t]}$ are the accumulated active sell and buy side trade quantity. 

The second indicator is the trade frequency imbalance rate, which calculates the proportion of active buy and sell trade numbers in the past period. Similar to the volume imbalance rate, when the number of active buy trades is greater than the number of active sell trades, this indicator predicts a higher probability of price increase. The specific expression is as follows:
\begin{equation}
    \mathrm{TIC}_t^{\delta}=\frac{\mathrm{TICS}_{[t-\delta, t]}-\mathrm{TICB}_{[t-\delta, t]}}{\mathrm{TICS}_{[t-\delta, t]}+\mathrm{TICB}_{[t-\delta, t]}},
\end{equation}
where $\mathrm{TICS}_{[t-\delta, t]}$ and $\mathrm{TICB}_{[t-\delta, t]}$ are the total number of active sell and buy side trades during the past time period. 

\begin{figure*}
  \begin{center}
  \includegraphics[width=.5\linewidth]{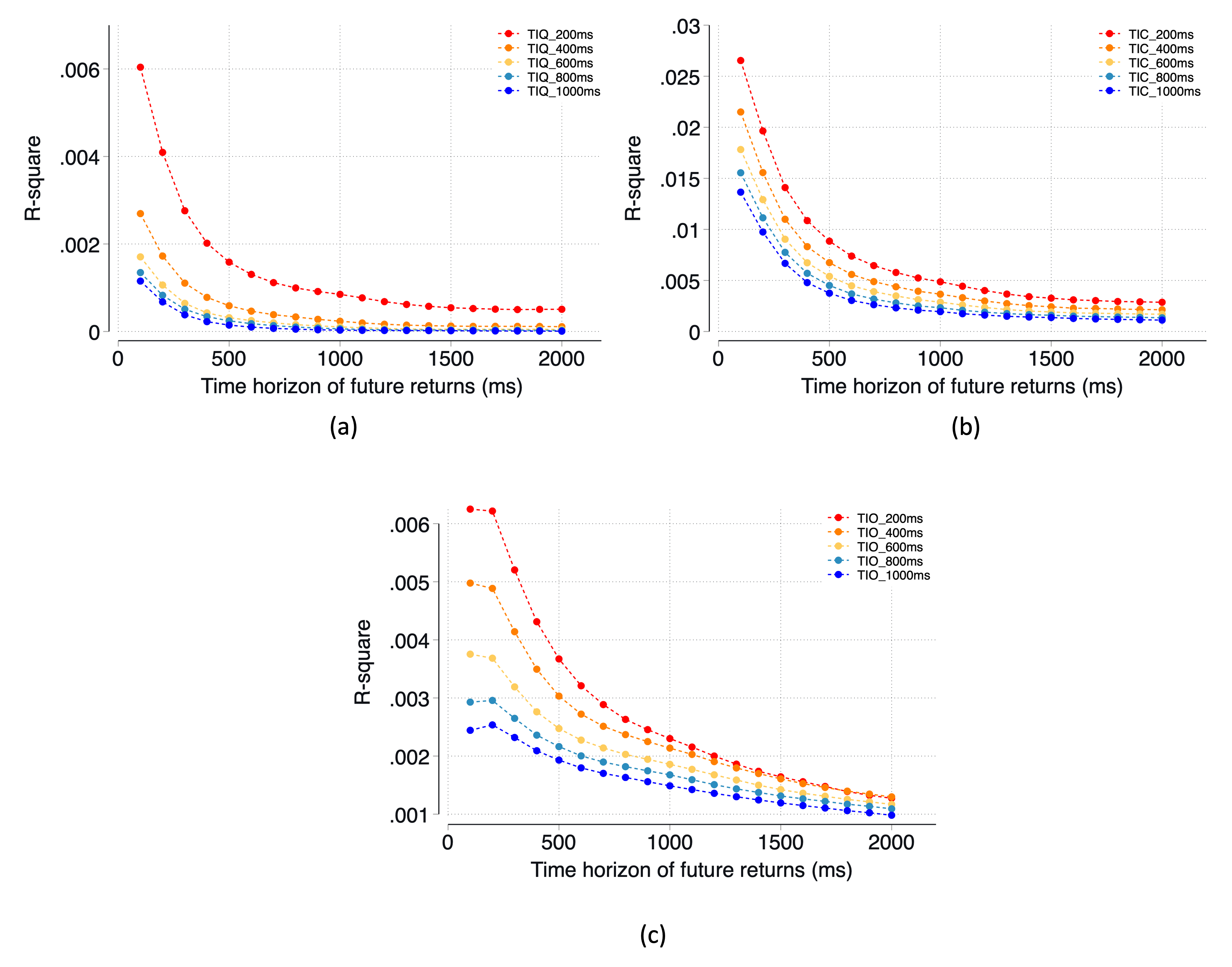}
  \end{center}
    \caption{\textbf{Comparison of $R^2$ values of trade flow imbalance indicators at different time horizons.} (a) trade imbalance of quantity (TIQ): the dashed lines represent the TIQ at different $\delta$ values. (b) trade imbalance of count (TIC): the dashed lines represent the TIC at different $\delta$ values. (c) trade imbalance of order submission and cancellation (TIO): the red lines represent the TIO in different time horizons.}
  \label{drl_mm_ti_r_square}
\end{figure*}

The third indicator is called the submission and cancellation rate of limit orders, which takes in account the amount of buy and sell limit orders submitted to the order book in the past period, as well as the amount of buy and sell limit orders canceled from the order book. Compared with indicators related to actual trades, this indicator is relatively conservative since when a limit order is submitted, it needs to wait on the order book and there is a chance of being executed or not being executed. Therefore, the intention to buy or sell is not as aggressive as the trade flow. However, this indicator also has a certain ability to judge the imbalance between buying and selling. The specific expression of this indicator is as follows:
\begin{equation}
\begin{split}
    \mathrm{TIOA}_{[t-\delta, t]} &= \mathrm{TIOA}^{submission}_{[t-\delta, t]} - \mathrm{TIOA}^{cancellation}_{[t-\delta, t]},\\
    \mathrm{TIOB}_{[t-\delta, t]} &= \mathrm{TIOB}^{submission}_{[t-\delta, t]} - \mathrm{TIOB}^{cancellation}_{[t-\delta, t]},\\
    \mathrm{TIO}_t^{\delta}&=\frac{\mathrm{TIOA}_{[t-\delta, t]}-\mathrm{TIOB}_{[t-\delta, t]}}{\mathrm{TIOA}_{[t-\delta, t]}+\mathrm{TIOB}_{[t-\delta, t]}},
\end{split}
\end{equation}
where $\mathrm{TIOA}^{submission}_{[t-\delta, t]}$ and $\mathrm{TIOB}^{submission}_{[t-\delta, t]}$ are the total quantity of ask limit orders and bid limit orders submitted to the order book during the past time period, $\mathrm{TIOA}^{cancellation}_{[t-\delta, t]}$ and $\mathrm{TIOB}^{cancellation}_{[t-\delta, t]}$ are the total quantity of ask limit orders and bid limit orders canceled from the order book.

We use linear regression to verify the predictive ability of the trade flow imbalance indicators, and the results are demonstrated in Figure \ref{drl_mm_ti_r_square}. From Figure \ref{drl_mm_ti_r_square} (a), it can be seen that TIQ's predictive ability is lower than OID or OIQ overall, and its predictive ability worsens as the length of look-back windows increases. The best performance is observed when TIQ's look-back window length is 200ms. Furthermore, similar to order book imbalance indicators, TIQ's predictive ability also decreases as the time horizon for future returns lengthens. Figure \ref{drl_mm_ti_r_square} (b) shows TIC's R-square metric, which demonstrates similar characteristics to TIQ, such as its predictive ability weakening as the look-back window length increases or the prediction time horizon increases. However, TIC's overall R-square is much larger than TIQ's. Figure \ref{drl_mm_ti_r_square} (c) displays TIO's R-square under different look-back windows, indicating that TIO's predictive ability is similar to TIQ's.

\begin{figure*}
  \begin{center}
  \includegraphics[width=0.5\linewidth]{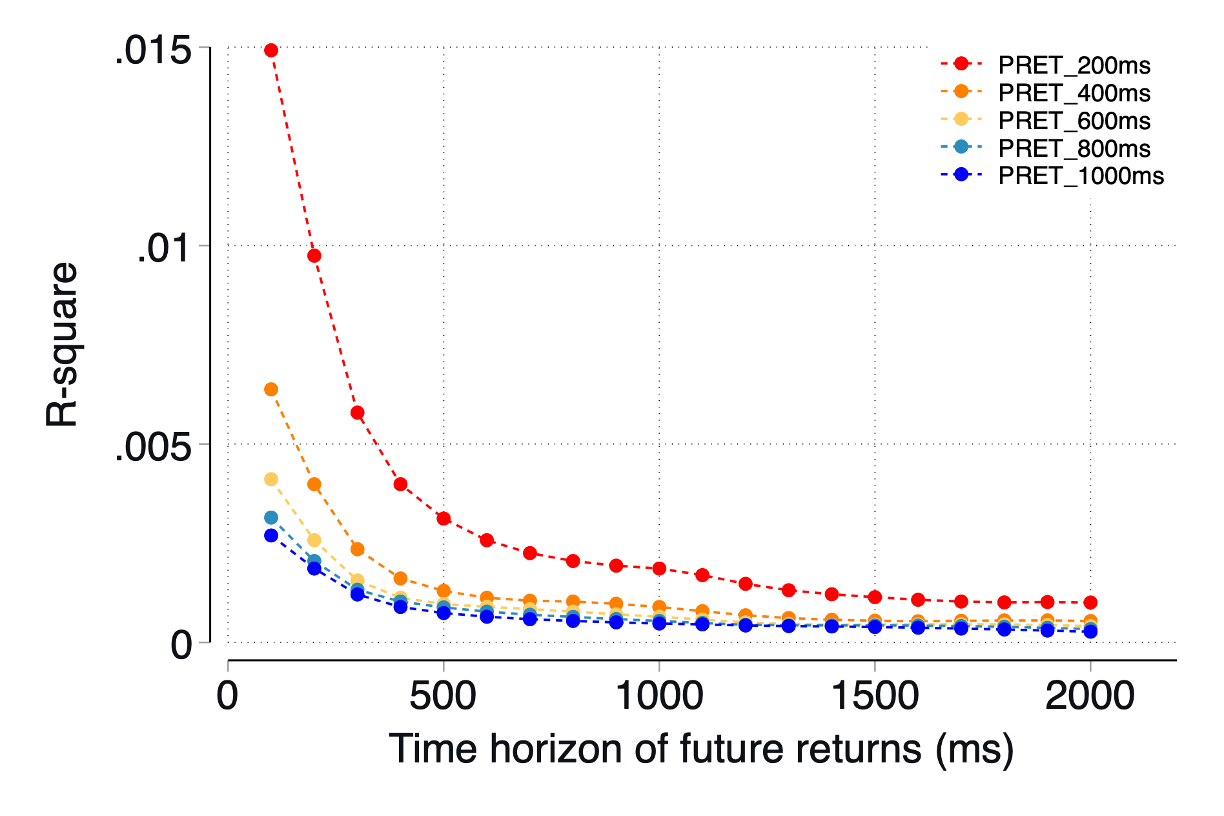}
  \end{center}
    \caption{\textbf{Comparison of $R^2$ values of post returns indicators at different time horizons with different $\delta$.}}
  \label{drl_mm_pret_r_square}
\end{figure*}

\subsection{Past Returns}
Past returns capture the historical price movements of the market, which also have a predictive ability for future returns in the regime of high-frequency trading. Especially when the market is in a trending stage, returns are more continuous, thus increasing the predictive power of past returns. To extract this indicator, we calculated the percentage change in price based on the time horizons we used in the above section. The definition is as follows:
\begin{equation}
    \operatorname{PRET}_t^{\delta}=\left(\frac{p_t^{mid}}{p_{t-\delta}^{mid}}-1\right),
\end{equation}
where $p_t^{mid}$ denotes the middle price at time $t$, $p_{t-\delta}^{mid}$ denotes the middle price at time $t-\delta$, and $\delta\in\left\{200ms, 400ms, 600ms, 800ms, 1000ms\right\}$.

Our evaluation in Figure \ref{drl_mm_pret_r_square} showed that past returns are a statistically significant and strong predictor of short-term price movements but have limited predictive power for longer-term trends. This feature is particularly useful for capturing short-term market movements and can be indicative of momentum and volatility.

\subsection{Discussion}
Based on the analysis results of the aforementioned factors, we can see that there are indeed various alpha signals in real market data, that is, historical signals have a certain predictive ability on future returns, but this predictive ability usually weakens as the prediction time lengthens. However, for high-frequency strategies, we focus more on short-term price prediction, because the holding time of high-frequency strategies is relatively short, which means that the strategy's net position will only be exposed to short-term price fluctuations, and is immune to long-term price changes. Therefore, these alpha signals can promote the performance of the high-frequency strategies. 

Specifically, the impact of alpha factor on the performance of high-frequency strategies is as follows: for high-frequency taker strategies, if the alpha signal predicts that the price is likely to rise significantly within the next 1 second, we should directly send market orders to buy the asset. If the signal indicates that the price will not rise anymore, we need to sell the asset immediately. The holding time of high-frequency taker strategies is generally within minutes, so it can be seen that the accuracy of the prediction signal is directly related to whether the high-frequency taker strategy can make a profit. Similarly, for high-frequency maker strategies, if the factor predicts that future prices will fall, we should adjust the price of the limit buy order downwards to avoid being filled by the takers who have stronger predictive ability, which can result in reverse selection. Therefore, adding predictive alpha signals is one of the main methods to reduce the risk of reverse selection for high-frequency maker strategies.

\section{Assembled Framework}\label{Assembled_Framework}
When it comes to embedding alpha factors with predictive capabilities into high-frequency market making strategies, we face several challenges: (1) different alpha factors have inconsistent data types, making it difficult to integrate the information contained within them; (2) most alpha factors are based on fixed-time or periodical slices, since this sampling method facilitates predictive analysis and statistical regression, but high-frequency market making strategies' actions are primarily triggered by event-driven data during execution, so integrating periodical signals with event-driven data is one of the challenges; (3) the logic design for strategy execution controlled by alpha factors is varied, so selecting or optimizing a suitable factor control logic is also a challenge.

To address the above challenges, we utilize a deep reinforcement learning model to merge market tick-level information with fixed time-sliced prediction information, and ultimately output event-driven actions. On the one hand, by merging these two different formats of data, it can reduce the risk of high-frequency market-making models being reverse-selected and effectively control inventory risk. On the other hand, through event-driven actions, market-making strategies can respond to sudden changes in the market in a timely manner, which also reduces the risk of reverse selection.

Market making strategies based on reinforcement learning models can extract information from input data through optimization methods, which provide us with a relatively simple and convenient information fusion method. Specifically, we only need to use the predicted information based on fixed time slices mentioned above as new features of observation to achieve the fusion of tick-level information and predicted information, which can help the deep reinforcement model output better actions. In the following parts, we verify the effectiveness of the assembled framework based on the performance of the market making strategies in real data scenarios.

\subsection{Experimental Setup}
\begin{figure*}
  \begin{center}
  \includegraphics[width=.5\linewidth]{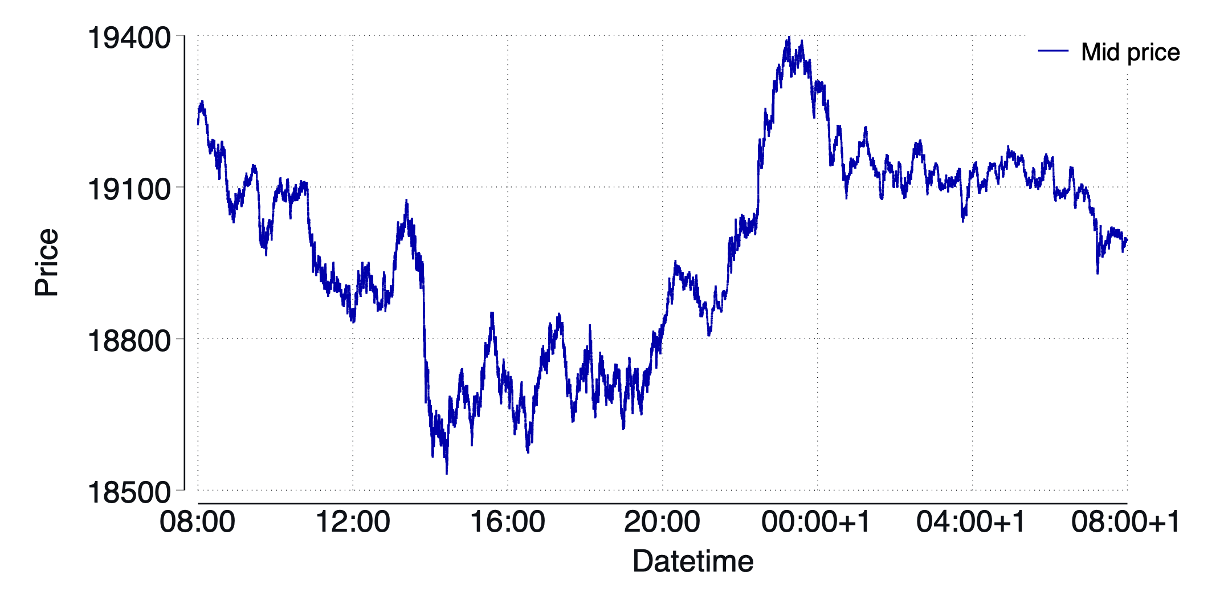}
  \end{center}
    \caption{\textbf{Real testing data}. This figure shows that the test data is from 8 am on September 23rd to September 24th, 2022, a total of 1 day.}
  \label{drl_mm_real_test_price}
\end{figure*}

Our experimental data is collected from the BTCUSD trading pair on the Coinbase exchange. The training data consists of real tick data from September 18th to September 22nd, 2022, total of 5 days, while the test data is from 8 am on September 23rd to September 24th, 2022, a total of 1 day, as showed in Figure \ref{drl_mm_real_test_price}. The training process is similar to that in the simulation part, where we randomly select tick data from the training dataset as the initial point in each episode, thereby increasing the randomness of the training set. In the testing process, we use the first tick data in the test dataset as the initial point and validate it in chronological order. In this experiment, we validated multiple models, including adaptive spread, double DQN, noisy net DQN, dueling DQN, and PPO models. We also compared the performance of models based on deep reinforcement learning algorithms with and without predictive information. 

\begin{figure*}
  \begin{center}
  \includegraphics[width=.5\linewidth]{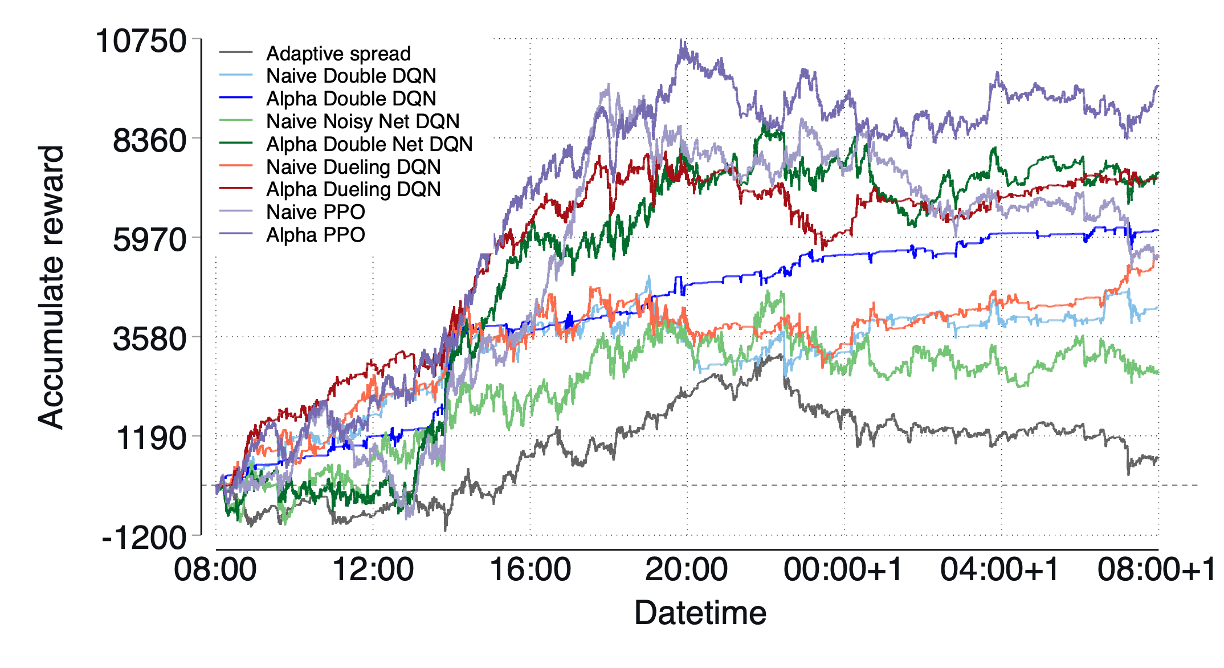}
  \end{center}
    \caption{\textbf{Performance of market making strategies based on different algorithms under real data testing}. From the chart, it can be seen that the cumulative return of the strategy based on Adaptive is the worst, while the one based on Alpha PPO performs the best.}
  \label{drl_mm_real_test_cumreward}
\end{figure*}

\subsection{Results}
\begin{figure*}
  \begin{center}
  \includegraphics[width=.5\linewidth]{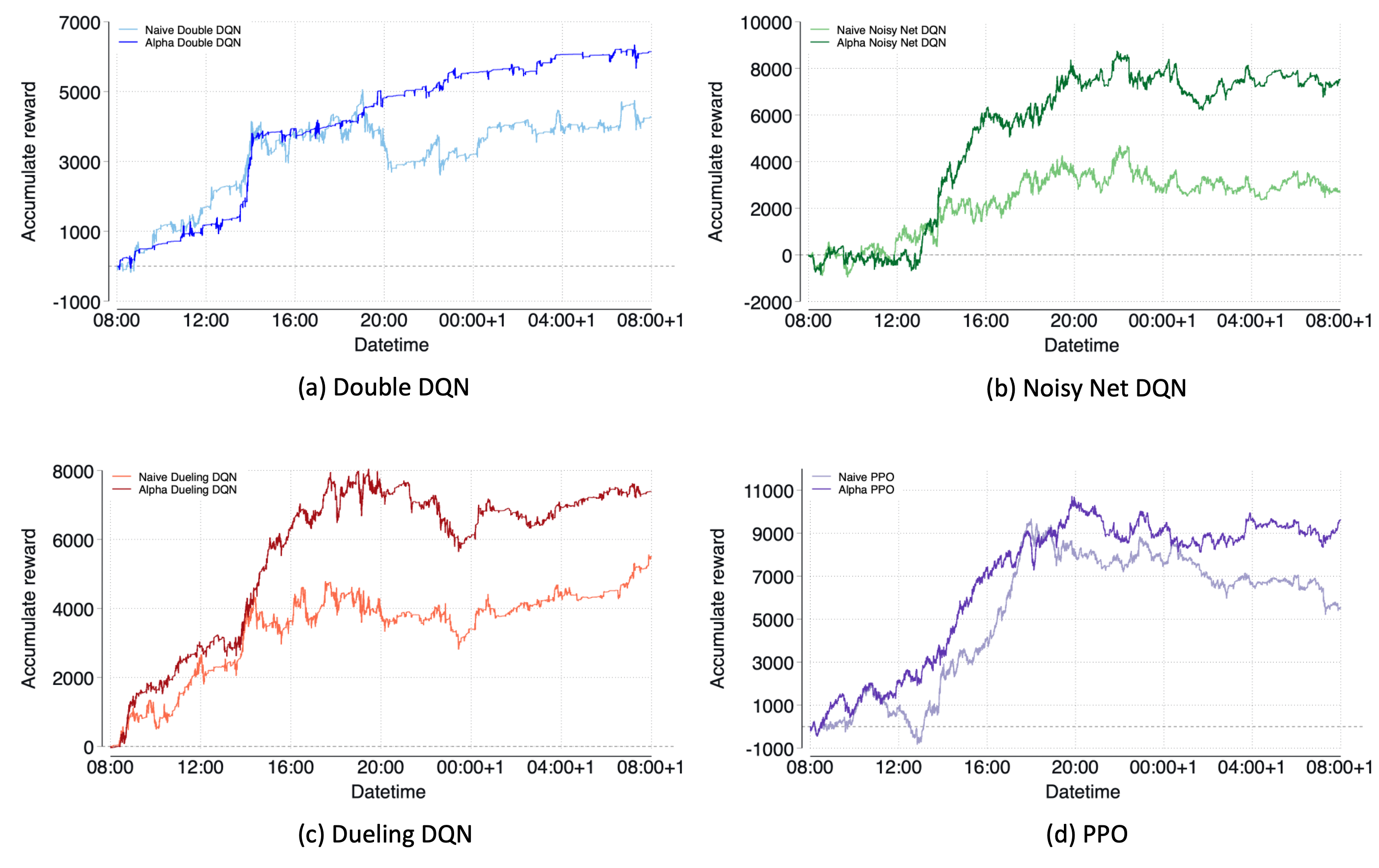}
  \end{center}
    \caption{\textbf{Performance comparison of different models between naive version and alpha version}. The figures below show the cumulative rewards of Double DQN, Noisy net DQN, Dueling DQN, and PPO in two different versions (naive version and alpha version). It can be observed that by incorporating alpha, the cumulative rewards of various models can be significantly improved.}
  \label{drl_mm_real_test_cumreward_detail}
\end{figure*}

\begin{figure*}
  \begin{center}
  \includegraphics[width=.5\linewidth]{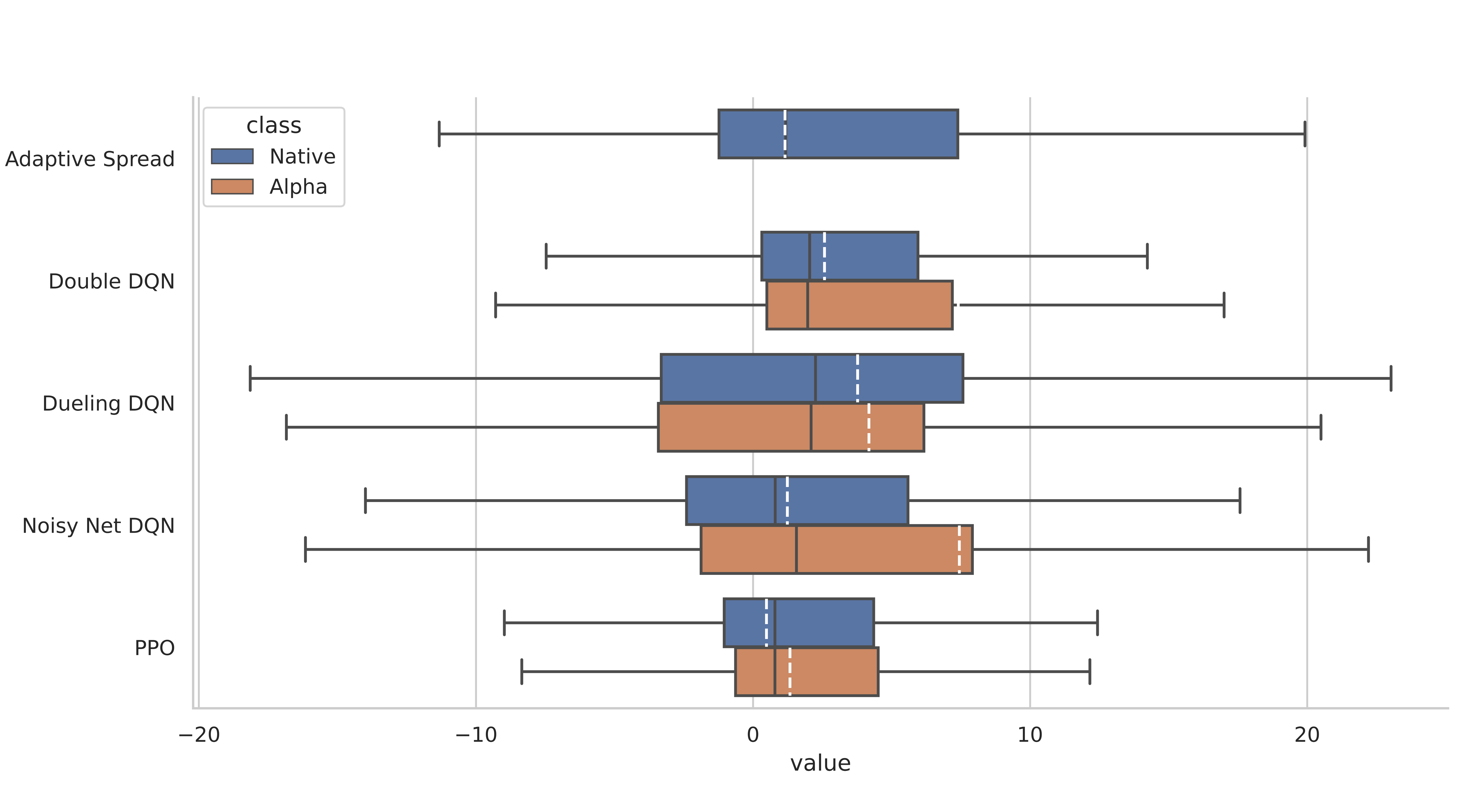}
  \end{center}
    \caption{\textbf{Boxplot of the reward of different models in each episode}. Comparing the simple model to its alpha version, we can see that the market-making strategy gets better rewards when alpha information is used. The reward's level of unpredictability stays mostly the same.}
  \label{drl_mm_real_test_game_reward_box}
\end{figure*}

Figure \ref{drl_mm_real_test_cumreward} shows the cumulative returns of strategies based on various algorithms in real data. It can be seen from the figure that the adaptive spread model, as the baseline, accumulated the maximum return of 3167 in the first 12 hours, but the return quickly declined to 666 in the following 12 hours. The strategy based on double DQN outperformed the adaptive spread strategy in the first 10 hours and achieved the highest return of 5065. Although the return declined to 3142 from 18:00 to 21:00, it steadily increased afterward and finally reached 4280. Compared with the double DQN model, the alpha double DQN model, which integrates predictive factors, as shown by the green line in the figure, did not decrease in return from 18:00 to 21:00. Therefore, the overall cumulative return was 6142, an increase of 1862 compared to the double DQN model. Similarly, we compared the noisy net DQN and alpha noisy net DQN strategies and found that the gap between them was larger than that between the two double DQN models, and the main difference came from the greater increase in return of alpha noisy net DQN from 13:00 to 20:00. Next, comparing the dueling DQN and alpha dueling DQN strategies, we can still see that the introduction of the alpha factor significantly improved the performance of the strategy. Finally, comparing the PPO and alpha PPO strategies, the figure shows that the PPO strategy obtained the maximum cumulative return near 18:00, and the return gradually declined to 5515 thereafter. On the other hand, the alpha PPO strategy achieved the maximum return of 10730 at 20:00 and the return rate was oscillating in the subsequent period, and the final return was 9610. Overall, in the real data scenario, methods based on deep reinforcement learning can improve returns on the baseline model, and the performance of such strategies can be further improved by introducing the alpha signal, demonstrating the effectiveness of the framework that integrates tick-level and periodical alpha signals through deep reinforcement learning algorithms. For comparison purposes, we separately drew the comparison between the basic naive version models, i.e., the version without integrating the alpha signal, and the alpha version models in Figure \ref{drl_mm_real_test_cumreward_detail}. From an episode perspective, Figure \ref{drl_mm_real_test_game_reward_box} shows the profit distribution of each episode in various strategies. By comparing the naive model with its corresponding alpha model, we can see that the average reward of the market-making strategy can be increased after integrating alpha information, and the volatility of the reward remains basically unchanged.

\subsection{Discussion}
This work proposes a unified reinforcement learning framework for market making strategies with latency, combining information from tick-level limit order book data and periodical data. A deep reinforcement learning model is developed with a tick-level environment and a latency-aware agent to learn tick-level information from event-driven limit order book data, while a feature extraction framework is designed to learn predictive information from periodical market data. The information from both tick-level and periodical data is embedded into the training phase of the DRL model to improve market making strategy performance.

\bibliographystyle{unsrtnat}
\bibliography{references}  






\end{document}